\documentclass[fp,twocolumn]{jpsj3}
\usepackage{txfonts}
\usepackage{bm}
\usepackage[dvipdfmx]{graphicx}
\usepackage[dvipdfmx]{color}
\usepackage{amsmath, amssymb}
\usepackage{braket}
\usepackage{color}
\usepackage{comment}

\title {Modeling and Physics of Multiferroic Perovskite Manganites}
\author{Masahito Mochizuki\thanks{masa{\_}mochizuki@waseda.jp}}
\inst{Department of Applied Physics, Waseda University, Okubo, Shinjuku-ku, Tokyo 169-8555, Japan}
\abst{A new type of multiferroicity was experimentally discovered in 2003 in a perovskite manganite TbMnO$_3$ where its ferroelectricity is induced by cycloidally ordered Mn spins. Susequently, such spin-cycloid multiferroic phase was also discovered in $R$MnO$_3$ with other rare-earth ions $R$=Dy, Eu$_{1-x}$Y$_x$, Tb$_{1-x}$Gd$_x$, etc. In this class of materials, the magnetism and ferroelectricity are inseparably coupled, and resulting strong magnetoelectric coupling enables us to control/manipulate the electricity (magnetism) by magnetic (electric) fields. Moreover, many interesting magnetoelectric phenomena due to their cross correlation have been discovered. In this article, we discuss a microscopic theoretical model for $R$MnO$_3$ constructed by taking into account their precise electronic and lattice structures and overview the theoretical works based on this model which elucidated rich magnetoelectric phenomena of $R$MnO$_3$. The perovskite manganites are not only the first-discovered spin-spiral multiferroic materials but also a typical class of materials that exhibits most of the magnetoelectric phenomena manifested in many other multiferroics. Therefore, the comprehensive understanding of $R$MnO$_3$ directly leads to the clarification of universal physics of magnetoelectric phenomena in multiferroic materials.}

\begin{document}
\maketitle
\section{Introduction}
There are materials called multiferroics. The original definition of the term ``multiferroics" is a class of materials in which multiple ferroic orders coexist~\cite{Schmid1994}. The ferroic orders include (anti)ferromagnetism, (anti)ferroelectrics, (anti)ferroelasticity, and any other (anti)ferroic orders of various degrees of freedom, e.g., helicity, vorticity, multipoles, toroidal moment, etc. Recently, the term is often used particularly for magnetoelectric materials, that is, materials in which magnetism and electricity coexist. The magnetism is not restricted to the (anti)ferromagnetism but is generalized to any other magnetic orders. In fact, many examples of magnetoelectric materials with coexisting magnetism and ferroelectricity had been known so far~\cite{Schmid1994}. However, in these old type magnetoelectric materials, the coexistence happens rather accidentally, and their coupling is relatively weak. Consequently, the magnetism and ferroelectricity are almost independent properties, and unique phenomena due to the coexistence could not be expected.

\begin{figure} [tb]
\begin{center}
\includegraphics[scale=1.0]{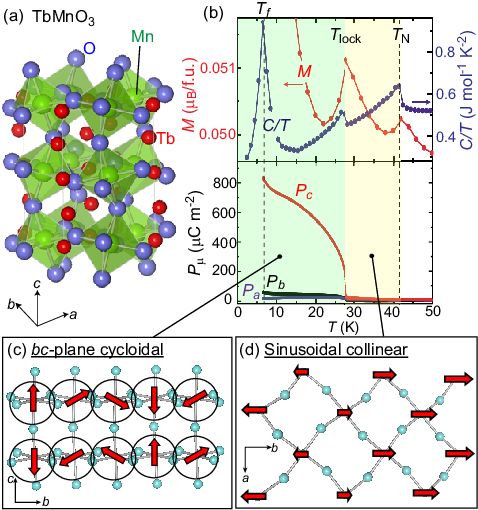}
\caption{(Color online) (a) Crystal structure of TbMnO$_3$. (b) Experimentally measured temperature-profiles of physical quantities~\cite{Kimura03a}. Magnetization $M$ and specific heat $C/T$ in the upper panel indicate successive three magnetic phase transitions. Electric polarization $\bm P$ in the lower panel exhibits increase of $P_c$ from the second transition point at $T_{\rm lock}$ indicating the magnetism-induced ferroelectricity. (c) Cycloidal Mn-spin order in the ferroelectric phase below $T_{\rm lock}$. (d) Sinusoidal collinear Mn-spin order in the paraelectric phase at intermediate temperatures between $T_{\rm N}$ and $T_{\rm lock}$. The third transition at $T_f$ corresponds to the ordering of Tb $f$-moments. Figure (b) is taken and modified from Ref.~\cite{Kimura03a} {\copyright} 2003 American Physical Society.}
\label{Fig01}
\end{center}
\end{figure}
In 2003, Kimura, Tokura and their coworkers discovered a material in which a special type of magnetic order induces a ferroelectric order~\cite{Kimura03a}. The material is the perovskite Mn oxide TbMnO$_3$ [Fig.~\ref{Fig01}(a)]. In this material, the ferroelectric polarization is induced by a spiral magnetic order of Mn spins, and, thereby, the magnetism and the ferroelectricity are strongly coupled. More specifically, it was experimentally revealed that TbMnO$_3$ exhibits successive three phase transitions with decreasing temperature [see upper panel of Fig.~\ref{Fig01}(b)], and the ferroelectricity appears at the second transition point, below which the ferroelectric polarization increases with decreasing temperature [see lower panel of Fig.~\ref{Fig01}(b)]. A subsequent neutron-scattering experiment revealed that a cycloidal order of Mn spins ($bc$-plane cycloidal order) shown in Fig.~\ref{Fig01}(c) is realized in the ferroelectric phase~\cite{Kenzelmann05}. On the other hand, the magnetic order in the paraelectric phase at intermediate temperatures is found to be a sinusoidal collinear order as shown in Fig.~\ref{Fig01}(d). These observations indicate that the ferroelectricity in TbMnO$_3$ is induced by the cycloidal order of Mn spins.

\begin{figure} [tb]
\begin{center}
\includegraphics[scale=1.0]{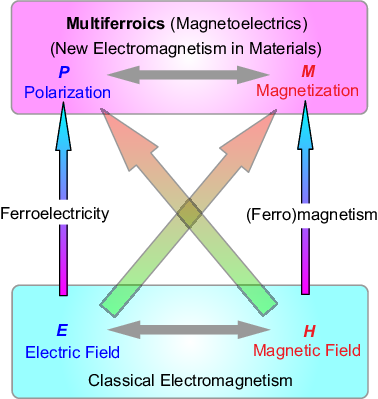}
\caption{(Color online) Cross correlation of magnetism and electricity in multiferroic materials with magnetism-induced ferroelectricity. The magnetoelectric coupling enables us to control and/or manipulate the electricity (magnetism) with magnetic (electric) fields.}
\label{Fig02}
\end{center}
\end{figure}
A revolutionary point of the discovery in Ref.~\cite{Kimura03a} is that a magnetic order breaks spatial inversion symmetry and induces a ferroelectric polarization, even though the original crystal structure has spatial inversion symmetry. This is in striking contrast to conventional multiferroics with accidentally coexisting magnetism and ferroelectricity. Taking advantage of the strong magnetoelectric coupling, we can control the ferroelectricity by magnetic fields and the magnetism by electric fields (Fig.~\ref{Fig02}). In fact, the magnetoelectric coupling in materials was originally predicted by Pierre Curie in 1894~\cite{PCurie1894}. However, before the discovery of magnetoelectricity in TbMnO$_3$, there had been no known material in which strong coupling between magnetism and ferroelectricity is realized. Hence, the discovery brought a breakthrough in the research on magnetoelectric coupling in multiferroics~\cite{Tokura06,Tokura07,SWCheong07,Khomskii09}.

As will be argued later, the spin-spiral-induced multiferroic phase has been discovered also in perovskite manganites with other rare-earth ions or their mixtures such as $R$=Dy, Eu$_{1-x}$Y$_x$, Gd$_{1-x}$Tb$_x$, etc. Moreover, it was discovered that the perovskite manganites with much smaller rare-earth ions such as Ho, Y, Lu, etc. host another magnetism-induced multiferroic phase with a different physical mechanism. Intensive studies have discovered a variety of interesting phenomena due to the strong magnetoelectric coupling in these phases. The perovskite manganite system has been recognized as a treasure box of the multiferroic phases and magnetoelectric phenomena. Therefore, it has been expected that clarification of their physics necessarily leads to comprehensive understanding of the magnetoelectric coupling and phenomena in multiferroic materials. 

However, their theoretical description is not easy because there inherently exist interplays of multiple degrees of freedom (i.e., charges, spins, orbitals, and lattices) and keen competitions among various magnetic interactions and anisotropies behind the multiferroic phenomena. Simplified theoretical models often fail to capture them. To investigate such a system, it is important to reproduce and capture the physical phenomena quantitatively using a precise microscopic model first. Then we should extract underlying physics by systematic analyses of the model. The physics thus clarified enables us to make reliable predictions for new phenomena. In the meanwhile, when we construct the model, the parameter values for magnetic interactions and magnetic anisotropies are calculated using the perturbation theory with respect to the spin-orbit interactions and/or transfer integrals. In such calculations, knowledge of electronic structures in the magnetic ions provided by the Tanabe-Sugano diagrams are indispensable~\cite{TanabeSugano1,TanabeSugano2,TanabeSugano3}. The epoch-making theoretical work achieved by Prof. Satoru Sugano and Prof. Yukito Tanabe 70 years ago is still vivid and plays vital roles in the development of the mutliferroics research, which is one of the most advanced fields of the modern science.

In this article, we overview previous theoretical studies based on a microscopic theoretical model for the perovskite manganites constructed by taking into account their electronic and lattice structures. The studies have revealed numerous physics and mechanisms behind the following magnetoelectric phenomena in the perovskite manganites,
\begin{itemize}
\item Polarization flop induced by the lattice distortion~\cite{Mochizuki09a,Mochizuki09b}
\item Multiferroic $E$-type antiferromagnetic phase with Peierls-type spin-lattice coupling~\cite{Mochizuki10a,Furukawa10,Mochizuki11a}
\item Magnetic-field-induced polarization switching and phase diagrams in magnetic field~\cite{Mochizuki10b,Tokunaga09,Matsubara15,Mochizuki15}
\item Multiferroic domain walls and giant magnetocapacitance effect~\cite{Kagawa09}
\item Electromagnon excitations activated by light electric field~\cite{Mochizuki10c,Mochizuki10d}
\item Predicted optical switching of spin chirality via intense electromagnon excitations~\cite{Mochizuki10d,Mochizuki11b}
\end{itemize}
The remained part of this article is organized as follows. In Sec.2, we introduce the spin-current model and the inverse Dzyaloshinskii-Moriya model as a physical mechanism of the ferroelectricity indued by spiral magnetism. In Sec.3, we discuss the properties and phase diagrams of the perovskite manganites. In particular, we argue several puzzles in the phase diagrams which cannot be explained by simple toy models. In Sec.4, we discuss various behaviors of ferroelectricity under application of magnetic field and related puzzling issues. In Sec.5, we construct a microscopic model of the perovskite manganites and argue a theoretical phase diagram obtained by this model. In Sec.6, we discuss the magnetoelectric phenomena in the dynamical regime. In particular, we argue the physical mechanism of magnon excitations activated by light electric field (called electromagnon) and predicted possible optical switching of spin chirality via intense electromagnon excitations. Section 7 is devoted to the conclusion.

\section{Spin-Induced Ferroelectricity}
Immediately after the discovery of the magnetism-induced ferroelectricity in TbMnO$_3$~\cite{Kimura03a}, a theory called spin-current model was proposed for its physical mechanism~\cite{Katsura05}, which revealed that mutually canted two spins on the adjacent magnetic ions with an in-between ligand ion can induce an electric polarization in the presence of the spin-orbit coupling. The cycloidal Mn-spin order observed in the subsequent neutron-scattering experiment confirmed the validity of this theory~\cite{Kenzelmann05}. Since then, the spin-current model has been a guiding principle for search for new materials, and numerous magnetism-induced multiferroic materials have indeed been discovered. In this section, we discuss theoretical models for the physical mechanism of the cycloidal-spin-induced ferroelectricity.

\subsection{Dzyaloshinskii-Moriya interaction}
\begin{figure} [tb]
\begin{center}
\includegraphics[scale=1.0]{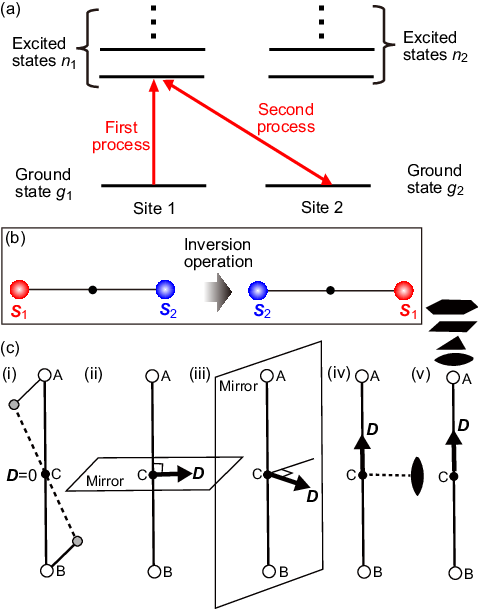}
\caption{(Color online) (a) Schematic diagram of the perturbation processes for the Dzyaloshinskii-Moriya interaction. (b) Space inversion operation for a bond connecting sites 1 and 2. (c) Rules for the presence or absence of nonzero Dzyaloshinskii-Moriya vector $\bm D$ and its orientation for a bond connecting two magnetic ions $A$ and $B$ with their midpoint C~\cite{Moriya60a,Magnetism1963}.}
\label{Fig03}
\end{center}
\end{figure}
For the magnetism-induced ferroelectricity in TbMnO$_3$, the Dzyaloshinskii-Moriya interaction~\cite{Dzyaloshinsky58,Moriya60a,Moriya60b} plays a substantial role. Here, we briefly explain properties of this interaction. The Dzyaloshinskii-Moriya interaction is antisymmetric exchange interaction working between two spins $\bm S_1$ and $\bm S_2$ on the adjacent lattice sites 1 and 2, which is given with their cross product $\bm S_1 \times \bm S_2$ as,
\begin{align}
\mathcal{H}_{\rm DM}=\bm D_{12} \cdot (\bm S_1 \times \bm S_2)
\label{eqn:DMI}
\end{align}
Here $\bm D$ is a real vector called Dzyaloshinskii-Moriya vector. This interaction originates from the perturbation processes with respect to the relativistic spin-orbit interaction and the exchange interaction [Fig.~\ref{Fig03}(a)]. The microscopic formula of the Dzyaloshinskii-Moriya vector is give by,
\begin{align}
\bm D_{12}&=-2i\lambda
\left[
\sum_{n_1}\frac{\braket{g_1|\bm L_1|n_1}}{E_{n_1}-E_{g_1}}J(n_1g_2,g_1g_2)
\right.
\notag \\
& \hspace{10mm} \left.
-\sum_{n_2}\frac{\braket{g_2|\bm L_2|n_2}}{E_{n_2}-E_{g_2}}J(g_1n_2,g_1g_2)
\right]
\label{eqn:DMV}
\end{align}
The first term describes the following perturbation process [see also Fig.~\ref{Fig03}(a)]. The electron on site 1 at the ground state $g_1$ is first excited to the excitation state $n_1$ by the spin-orbit interaction as the first process, and then the excited electron returns back to the ground state $g_1$ through interacting with the ground state $g_2$ on site 2 via the exchange interaction $J(n_1g_2,g_1g_2)$ as the second process. On the contrary, the second term in Eq.~(\ref{eqn:DMV}) describes another perturbation process starting from the ground state $g_2$ on site 2. The magnitude of $|\bm D_{12}|$ is the order of $D \approx \lambda J/\Delta E$ where $J$ and $\lambda$ are the coupling constants of the (anti)ferromagnetic exchange interaction and the spin-orbit coupling, respectively, and $\Delta E$ is the energy-level difference between the ground state and the excited states.

Importantly, the sign of the Dzyaloshinskii-Moriya vector is reversed when the sites 1 and 2 are exchanged, i.e., $\bm D_{12}=-\bm D_{21}$. This fact results in another important property, that is, the Dzyaloshinskii-Moriya vector becomes zero when the midpoint between the site 1 and site 2 is the inversion center. Specifically, in the presence of spatial inversion symmetry, $\bm D_{12}$ and $\bm D_{21}$ must be zero in order to satisfy both of the following relations, $\bm D_{12} \cdot (\bm S_1 \times \bm S_2)=\bm D_{12} \cdot (\bm S_2 \times \bm S_1)$ due to the inversion symmetry and $\bm D_{12} \cdot (\bm S_1 \times \bm S_2)=-\bm D_{12} \cdot (\bm S_2 \times \bm S_1)$ due to the antisymmetric nature of cross vector products [see [Fig.~\ref{Fig03}(b)]. According to this property, the following rules hold for Dzyaloshinskii-Moriya vectors when we consider a bond connecting two magnetic ions A and B with C as their midpoint [Fig.~\ref{Fig03}(c)], that is, (i) when there is inversion symmetry at C, $\bm D$=0, (ii) when there is a mirror plane perpendicular to A-B at C, $\bm D$ is within the mirror plane ($\bm D$$\parallel$A-B), (iii) when there is a mirror plane containing A and B, $\bm D$ is perpendicular to the mirror plane, (iv) when there is a $2$-fold rotation axis perpendicular to A-B passing through C, $\bm D$ is perpendicular to the rotation axis, and (v) when there is an $n$-fold rotation axis ($n \ge 2$) along A-B, $\bm D$ is parallel to A-B~\cite{Moriya60a,Magnetism1963}.

\subsection{Spin-current model}
\begin{figure*} [tb]
\begin{center}
\includegraphics[scale=1.0]{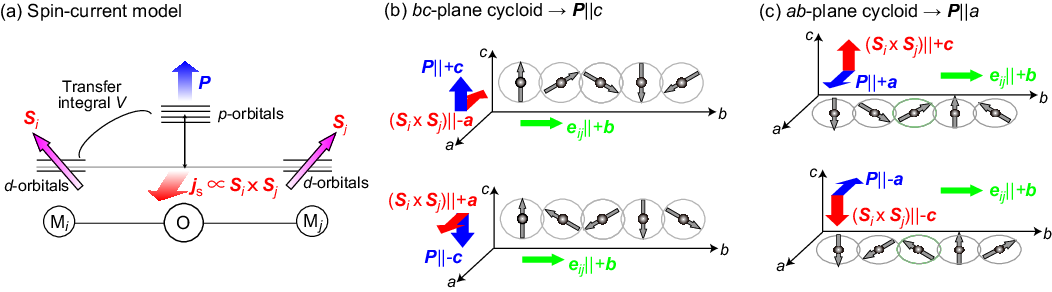}
\caption{(Color online) (a) Schematics of the spin-current model as a physical mechanism of the magnetism-induced electric polarization in Eq.~(\ref{eqn:KNB}). (b),~(c) Relationship between the cycloidal spin order and the induced ferroelectric polarization described by the spin-current model. When the spin cycloid propagates along the $b$ axis (i.e., $\bm e_{ij}$$\parallel$$\bm b$), the $bc$-plane ($ab$-plane) spin cycloid with $\bm S_i$$\times$$\bm S_j$$\parallel$$\bm a$ ($\bm S_i$$\times$$\bm S_j$$\parallel$$\bm c$) induces the ferroelectric polarization $\bm P$$\parallel$$\bm c$ ($\bm P$$\parallel$$\bm a$). Equation~(\ref{eqn:KNB}) also indicates that the sign of $\bm P$ is reversed when the rotation sense of the spin cycloid is reversed.}
\label{Fig04}
\end{center}
\end{figure*}
The magnetism-induced ferroelectricity in TbMnO$_3$ has two dominant contributions. One is the asymmetric charge distribution, the other is the uniform lattice displacement. Both of these effects are caused by the spin-orbit interaction or the Dzyaloshinskii-Moriya interaction in the presence of canted spin pairs $\bm S_i$ and $\bm S_j$ on the adjacent sites $i$ and $j$. Interestingly, both contributions are represented by the same formula, 
\begin{equation}
\bm p_{ij}=A\bm e_{ij} \times (\bm S_i \times \bm S_j).
\label{eqn:KNB}
\end{equation}
This equation indicates that two mutually canted spins $\bm S_i$ and $\bm S_j$ induce a local electric polarization $\bm p_{ij}$ proportional to their cross product $\bm S_i$$\times$$\bm S_j$. Here $\bm e_{ij}$ is the unit direction vector connecting the two sites $i$ and $j$, and $A$ is the coupling constant determined by the spin-orbit coupling, the exchange interaction, and other relevant interactions. This formula was first proposed by Katsura, Nagaosa, and Balatzky in 2005~\cite{Katsura05}. They derived this formula by the perturbation theory with respect to transfer integrals through considering two transition-metal ions with $d$-orbitals split by the spin-orbit interaction, the in-between ligand ion with $p$ orbitals, and the repulsive Coulomb interaction among electrons on the $d$ orbitals [Fig.~\ref{Fig04}(a)]. The derivation is based on the Tanabe-Sugano diagrams~\cite{TanabeSugano1,TanabeSugano2,TanabeSugano3}, which provides knowledge of the electronic structure.

In fact, Eq.~(\ref{eqn:KNB}) can be derived also from a spin-lattice coupled model with the Dzyaloshinskii-Moriya interaction~\cite{Sergienko06a} and a phenomenological theory~\cite{Mostovoy06}. The mechanism described by this equation has various names, e.g., (1) spin-current mechanism, (2) Katsura-Nagaosa-Balatzky mechanism, (3) inverse Dzyaloshinskii-Moriya mechanism, (4) antisymmetric magnetostriction mechanism, and (5) S cross S mechanism. The names (1) and (2) imply a mechanism based on the asymmetric charge distribution due to the spin-orbit interaction, while the names (3) and (4) imply a mechanism based on the uniform lattice displacement due to the Dzyaloshinskii-Moriya interaction as an origin of the ferroelectric polarizations. On the contrary, the name (5) is a general name used for both mechanisms.

Equation~(\ref{eqn:KNB}) indicates that cycloidal magnetism as a series of canted spins can induce the ferroelectricity with uniform electric polarizations. In the introduction section, we mentioned that the neutron-scattering experiment performed by Kenzelmann and coworkers observed a cycloidal order of Mn spins in the multiferroic phase of TbMnO$_3$~\cite{Kenzelmann05}. Here we should note that Katsura, Nagaosa, and Balatzky proposed the spin-current mechanism prior to this experiment. It is surprising that theorists established a prescient theory on the magnetism-induced ferroelectricity with no prior knowledge of the magnetic structure in TbMnO$_3$.

According to the form of Eq.~(\ref{eqn:KNB}) , the orientation of ferroelectric polarization is governed by the orientation of spin cycloidal plane. As seen in Fig.~\ref{Fig04}(b), when the cycloidal spin order propagate along the $b$ axis (i.e., $\bm e_{ij}$$\parallel$$\bm b$), the $bc$-plane spin cycloid with $\bm S_i$$\times$$\bm S_j$$\parallel$$\bm a$ is expected to induce ferroelectric polarization $\bm P$$\parallel$$\bm c$. On the contrary, as seen in Fig.~\ref{Fig04}(c), the $ab$-plane spin cycloid with $\bm S_i$$\times$$\bm S_j$$\parallel$$\bm c$ is expected to induce $\bm P$$\parallel$$\bm a$. This prediction was indeed validated by subsequent neutron-scattering experiments~\cite{Arima06,Yamasaki08}.

The orientation of the spin cycloidal plane is usually determined by competition among the magnetic anisotropies and the Dzyaloshinskii-Moriya interactions. For a series of the $R$MnO$_3$ compounds with $R$ being a rare-earth ion, not only TbMnO$_3$ but also DyMnO$_3$ exhibit the $bc$-plane spin cycloid at the ground state, while Eu $_{1-x}$Y$_x$MnO$_3$ ($x$$\sim$0.5) exhibits the $ab$-plane spin cycloid. Consequently, TbMnO$_3$ and DyMnO$_3$ have ferroelectric polarization $\bm P$$\parallel$$\bm c$, while Eu $_{1-x}$Y$_x$MnO$_3$ ($x$$\sim$0.5) has $\bm P$$\parallel$$\bm a$. In the meanwhile, the magnitudes of $\bm P$ in these compounds are, at most, $P_c$$\sim$500$\mu$C/m$^2$ for TbMnO$_3$~\cite{Kimura03a}, $P_c$$\sim$2000$\mu$C/m$^2$ for DyMnO$_3$~\cite{Goto04}, and $P_a$$\sim$1000$\mu$C/m$^2$ for Eu$_{1-x}$Y$_x$MnO$_3$ ($x$$\sim$0.5)~\cite{Yamasaki07b}.

Equation~(\ref{eqn:KNB}) further indicates that the spin chirality $\bm C$=$\frac{1}{N}\sum_{\langle i,j\rangle}\bm S_i \times \bm S_j$ or the rotation sense of the Mn-spin cycloid should be reversed when the ferroelectric polarization is reversed by application of an electric field. Indeed, a neutron-scattering experiment revealed that the spin chirality is reversed upon the polarization reversal from $\bm P$$\parallel$$+\bm c$ to $\bm P$$\parallel$$-\bm c$ in TbMnO$_3$~\cite{Yamasaki07a}.

\subsection{Inverse Dzyaloshinskii-Moriya mechanism}
\begin{figure} [tb]
\begin{center}
\includegraphics[scale=0.5]{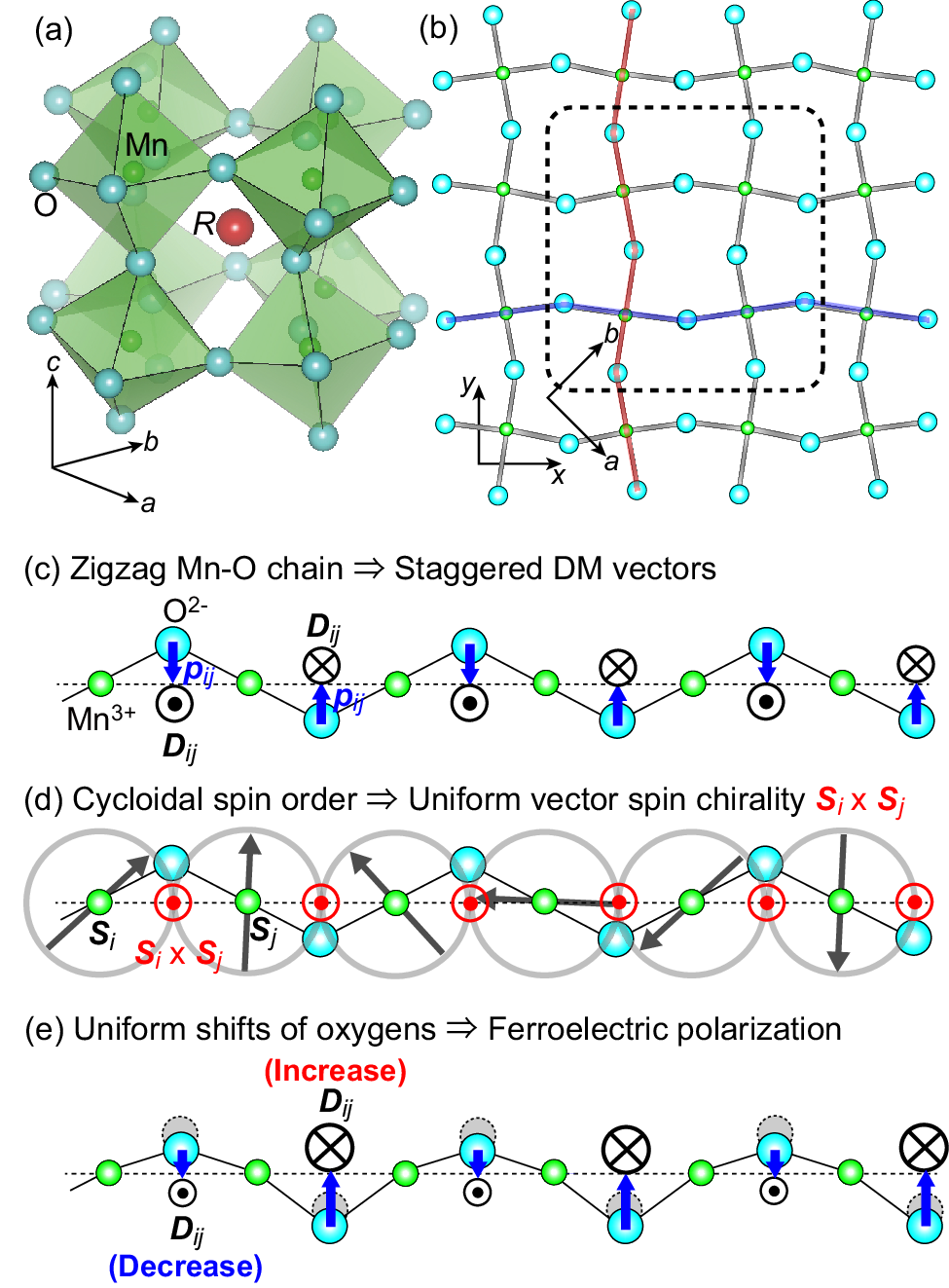}
\caption{(Color online) (a) Distorted perovskite structure with mutually tilted MnO$_6$ octahedra called GdFeO$_3$-type structure. (b) Zigzag Mn-O chains along the $x$ and $y$ axes in the $ab$ plane. (c)-(e) Step-by-step explanation of the inverse Dzyaloshinskii-Moriya mechanism as a physical origin of the spin-cycloid-induced ferroelectricity. (c) Staggered local electric polarizations $\bm p_{ij}$ and alternate Dzyaloshinskii-Moriya vectors $\bm D_{ij}$ on the zigzag Mn-O chain. (d) Cycloidal spin order with uniform local vector spin chirality $\bm S_i \times \bm S_j$. (e) Uniform shifts of the oxygen ions on the zigzag Mn-O chain which occur to enhance (reduce) the energy gain (cost) associated with the Dzyaloshinskii-Moriya interaction through increase (decrease) the magnitude of $\bm D_{ij}$ on the Mn-O-Mn bonds with $\bm D_{ij}$ antiparallel (parallel) to $\bm S_i \times \bm S_j$.}
\label{Fig05}
\end{center}
\end{figure}
Theoretical studies based on the first-principles calculations revealed that the contribution from the uniform lattice displacement is dominant as compared to the contribution from the asymmetric charge distribution for the ferroelectricity in TbMnO$_3$~\cite{HJXiang08,Malashevich08}. Here we argue how the cycloidal order of Mn spins can induce the uniform lattice displacement in TbMnO$_3$ via the inverse Dzyaloshinskii-Moriya mechanism.

The perovskite structure of TbMnO$_3$ is accompanied by a significant lattice distortion called GdFeO$_3$-type distortion associated with alternate tilting of the MnO$_6$ octahedra [Fig.~\ref{Fig05}(a)]. In this structure, the Mn-O-Mn bonds are no longer straight but form zigzag chains. The zigzag MnO-chains run along the $x$ and $y$ directions in the $ab$ plane as shown in Fig.~\ref{Fig05}(b). A nonzero Dzyaloshinskii-Moriya vector exists at the midpoint of the adjacent Mn ions for each bended Mn-O-Mn bond according to the rules argued above [see Fig.~\ref{Fig03}(c)-(iii)]. The vectors $\bm D_{ij}$ are aligned alternately on the zigzag bonds, and they have the same magnitude originally [Fig.~\ref{Fig05}(c)]. Note that there are staggered local electric polarizations $\bm p_{ij}$ on the zigzag Mn-O chain, each of which points from the negatively charged oxygen ion O$^{2-}$ to the midpoint of the positively charged adjacent two Mn $^{3+}$ ions on the Mn-O-Mn bond. Because they have the same magnitude but opposite signs, they cancel each other. 

We consider what happens if a cycloidal spin order is introduced in this situation. The cycloidally rotating spins have uniform local vector spin chirality $\bm S_i$$\times$$\bm S_j$ perpendicular to the cycloidal plane [Fig.~\ref{Fig05}(d)]. The Dzyaloshinskii-Moriya interaction energetically favors the antiparallel configuration of the Dzyaloshinskii-Moriya vector $\bm D_{ij}$ and the vector spin chirality $\bm S_i$$\times$$\bm S_j$ on each Mn-O-Mn bond. Consequently, the Mn-O-Mn bonds on which $\bm D_{ij}$ and $\bm p_{ij}$ are parallel decrease the extent of bond bending (increase the bond angle) through a spontaneous shift of the oxygen ion so as to reduce the cost of energy associated with the Dzyaloshinskii-Moriya interaction [Fig.~\ref{Fig05}(e)]. On the contrary, on the Mn-O-Mn bonds with antiparallel $\bm D_{ij}$ and $\bm p_{ij}$, a spontaneous shift of the oxygen ion occurs to enhance the energy gain of the Dzyaloshinskii-Moriya interaction through decreasing the bond angle. Importantly, these oxygen shifts occur in the same direction which give rise to a uniform component of the electric polarizations and resulting ferroelectricity.

\section{Properties of the Perovskite Manganites}
The spin-current model or the inverse Dzyaloshinskii-Moriya mechanism represented by Eq.~(\ref{eqn:KNB}) has provided a clear guideline to the search for new multiferroic materials, that is ``Look for materials with a cycloidal (or transverse spiral) magnetism!", which led to a dramatic progress in the subsequent research. In the intensive search for new materials based on this guideline, many new spin-induced multiferroic materials have been discovered, which include not only the Mn-perovskite family $R$MnO$_3$ ($R$=Dy, Eu$_{1-x}$Y$_x$, Gd$_{1-x}$Tb$_x$, etc) but also, for examples, Ba$_{0.5}$Sr$_{1.5}$Zn$_2$Fe$_{12}$O$_{22}$~\cite{Kimura05b}, Ni$_3$V$_2$O$_8$~\cite{Lawes05}, CoCr$_2$O$_4$~\cite{Yamasaki06}, MnWO$_4$~\cite{Taniguchi06,Heyer06,Arkenbout06}, LiCu$_2$O$_2$~\cite{SPark07,Seki08}, CuO~\cite{Kimura08} and many others. These materials have also turned out to exhibit interesting magnetoelectric phenomena as reviewed in Refs.~\cite{Kimura07,Seki10}.

\begin{figure} [tb]
\begin{center}
\includegraphics[scale=0.5]{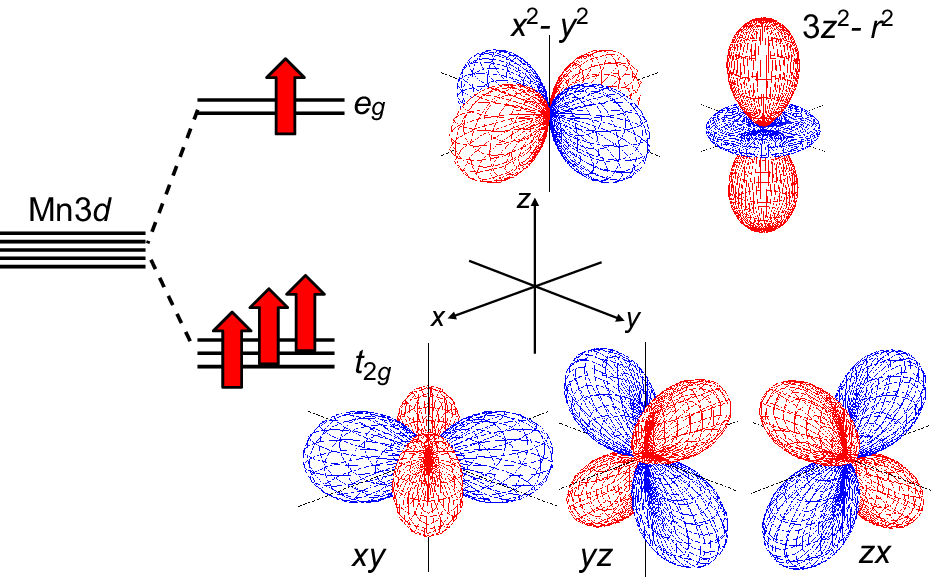}
\caption{(Color online) Electronic structure of Mn$^{3+}$ ion in $R$MnO$_3$ where $S$=2 high-spin state with $t_{2g}^3$$e_g^1$ electron configuration is realized due to the octahedral crystal field of ligand oxygens and the strong Hund's-rule coupling.}
\label{Fig06}
\end{center}
\end{figure}
The first discovered TbMnO$_3$ and its family $R$MnO$_3$ with $R$=Dy, Eu$_{1-x}$Y$_x$ and Gd$_{1-x}$Tb$_x$ are typical materials of cycloidal-spin multiferroics. Moreover, it was discovered that $R$MnO$_3$ compounds with smaller rare-earth ions ($R$=Ho, Er, Tm ,Yb, Lu, and Y$_{1-y}$Lu$_y$, etc.) also exhibit a magnetism-induced ferroelectric phase~\cite{Ishiwata10}, and its physical mechanism as well as its magnetic order are totally different from those in the cycloidal-spin multiferroics. This means that the $R$MnO$_3$ family is a typical class of multiferroic materials with magnetism-induced ferroelectricity in a broad sense. Therefore, we expect that clarification of physics of the $R$MnO$_3$ family will necessarily lead to the comprehensive understanding of rich magnetoelectric phenomena manifested by the multiferroic materials. In this section, we discuss fundamental properties and phase diagrams of $R$MnO$_3$. 

\subsection{Fundamental properties}
The crystal structure of $R$MnO$_3$ is a distorted perovskite structure with $Pbnm$ symmetry composed of mutually tilted MnO$_6$ octahedra as shown in Fig.~\ref{Fig05}(a). This lattice structure (lattice distortion) is called GdFeO$_3$-type structure (distortion). The magnitude of the lattice distortion depends on the size of the rare-earth ion. The Mn-O-Mn bond angle is reduced more significantly, and the lattice is more significantly distorted with a smaller rare-earth ion. The five-fold Mn $3d$-orbitals are split into lower-lying three-fold $t_{2g}$ orbitals and higher-lying two-fold $e_g$ orbitals by the octahedral crystal field of oxygen ions. Each Mn $^{3+}$ ion has a $t_{2g}^3$$e_g^1$ electron configuration where three $t_{2g}$ orbitals and one $e_g$ orbital are occupied, and a high-spin state with $S$=2 is realized by the Hund's-rule coupling as shown in Fig.~\ref{Fig06}.

\begin{figure} [tb]
\begin{center}
\includegraphics[scale=1.0]{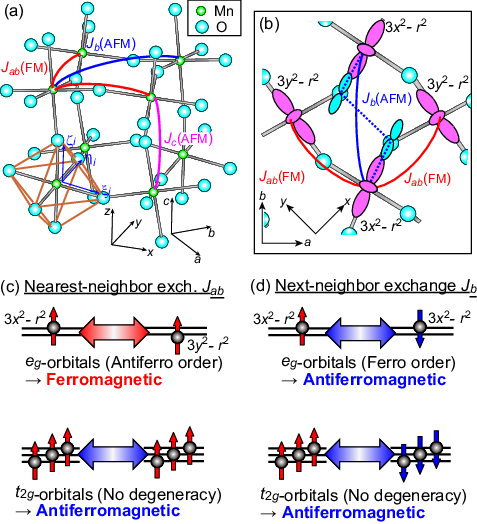}
\caption{(Color online) (a) Exchange interactions among Mn spins in $R$MnO$_3$ and local coordinate axes, $\xi_i$, $\eta_i$, and $\zeta_i$, attached to the tilted MnO$_6$ octahedron. (b) Exchange path for the in-plane next-neighbor antiferromagnetic exchange interaction $J_b$ along the $b$ axis (dotted line) and staggered orbital ordering of the occupied $e_g$ orbitals. (c),~(d) Contributions from the $e_g$-orbital and $t_{2g}$-orbital channels to the in-plane nearest-neighbor ferromagnetic exchange interaction $J_{ab}$ and the in-plane next-neighbor antiferromagnetic exchange interaction $J_b$. For $J_{ab}$, the staggered $e_g$ orbitals give a ferromagnetic contribution, whereas the $t_{2g}$ orbitals occupied by three electrons give an antiferromagnetic contribution. A partial cancelation of these opposite contributions results in the weakly ferromagnetic $J_{ab}$. For $J_b$, both $e_g$ and $t_{2g}$ orbitals give antiferromagnetic contributions, which results in the relatively strong antiferromagnetic interaction $J_b$.}
\label{Fig07}
\end{center}
\end{figure}
The cycloidal spin orders in TbMnO$_3$ and DyMnO$_3$ are realized by frustration between the nearest-neighbor ferromagnetic exchange interaction $J_{ab}$ and the next-neighbor antiferromagnetic exchange interaction $J_b$ in the $ab$ plane [see Figs.~\ref{Fig07}(a) and (b)]. The next-neighbor interaction $J_b$ originates from an exchange path through two oxygen 2$p$ orbitals as shown in Fig.~\ref{Fig07}(b)~\cite{Kimura03b}. When the GdFeO$_3$-type lattice distortion is more significant with larger tilting of the MnO$_6$ octahedra, these two oxygens become closer and consequently $ J_b$ becomes larger. Therefore, $J_b$ is a variable scaled to the extent of the GdFeO$_3$-type distortion or the ionic radius of the rare-earth ion.

Usually, the next-neighbor exchange interactions should be very weak in the perovskite compounds as compared to the nearest-neighbor exchange interactions and, hence, rarely play a major role. However, in the present $R$MnO$_3$ system, the nearest-neighbor interaction $J_{ab}$ is very weak for a certain reason explained below, and, hence, the next-neighbor interaction $J_b$ attains relative significance and plays an important role. This is a unique property of the $R$MnO$_3$ system due to its $t_{2g}^3$$e_g^1$ electron configuration of the Mn$^{3+}$ ions~\cite{Mochizuki09a}. 

There are two channels which give opposite contributions to the nearest-neighbor interaction $J_{ab}$. One is the $t_{2g}$-orbital channel between the $S$=3/2 spins, the other is the $e_g$ orbital channel between the $S$=1/2 spins. The former channel gives an antiferromagnetic contribution because the $t_{2g}$ orbitals are occupied by three electrons, while the latter channel gives a ferromagnetic contribution because of the antiferro orbital ordering~\cite{Roth66,KK72,KK73,Inagaki73,Cyrot73,Cyrot75} with staggered occupations of $d_{3x^2-r^2}$ and $d_{3y^2-r^2}$ orbitals. These opposite contributions partially cancel each other, resulting in the weak ferromagnetic interaction $J_{ab}$. On the contrary, the next-neighbor interaction $J_b$ tends to be relatively strong, since both $t_{2g}$-orbital channel and the $e_g$-orbital channel cooperatively contribute to this antiferromagnetic interaction.

\begin{figure} [tb]
\begin{center}
\includegraphics[scale=1.0]{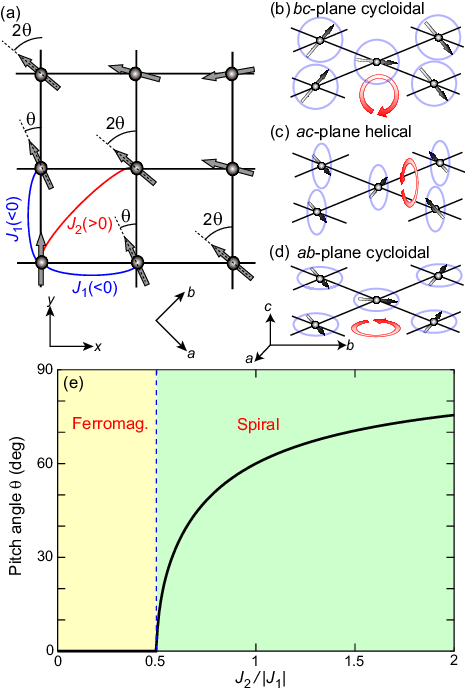}
\caption{(Color online) (a) Schematics of the $J_1$-$J_2$ classical Heisenberg model on the square lattice in Eq.~(\ref{eqn:J1J2M}) and the spiral spin order. (b)-(d) Spiral spin orders with different spiral plane. (e) Phase diagram of the model. A spiral spin phase with a pitch angle of $\theta=\cos^{-1}[|J_1|/(2J_2)]$ appears when $J_2/|J_1|>0.5$, whereas a ferromagnetic phase with $\theta=0$ appears when $J_2/|J_1|\leq0.5$.}
\label{Fig08}
\end{center}
\end{figure}
The frustration-induced spiral magnetism can be reproduced by a simple $J_1$-$J_2$ classical Heisenberg model on the square lattice [see Fig.~\ref{Fig08}(a)], which is given by,
\begin{align}
\mathcal{H}_{J_1J_2}&=
J_1\sum_{< i,j >}\bm S_i \cdot \bm S_j +J_2\sum_{\ll i,j \gg}\bm S_i \cdot \bm S_j 
+K \sum_iS_{\mu i}^2.
\label{eqn:J1J2M}
\end{align}
Here $J_1(<0)$ is set to be ferromagnetic, while $J_2(>0)$ is set to be antiferromagnetic. The third term with $K>0$ describes the easy-plane type magnetic anisotropy. This anisotropy determines the rotation plane of the spiral spins, that is, the $bc$-plane cycloidal, the $ac$-plane helical, and the $ab$-plane cycloidal spin orders are stabilized for $\mu$=$a$, $b$ and $c$, respectively [Figs.~\ref{Fig08}(b)-(d)]. Assuming a spiral spin configuration with an equivalent pitch angle $\theta$, the energy per site can be given as a function of $\theta$ in the form,
\begin{align}
E(\theta)/S^2=2J_1\cos\theta + J_2\cos(2\theta).
\end{align}
The saddle-point equation $dE(\theta)/d\theta=0$ tells us that a spiral spin phase with a pitch angle of $\theta=\cos^{-1}[|J_1|/(2J_2)]$ is realized when $J_2/|J_1|>0.5$. On the other hand, a ferromagnetic phase with $\theta=0$ appears when $J_2/|J_1|\leq0.5$. The phase diagram of this model is obtained as a function of $J_2$ [Fig.~\ref{Fig08}(e)]. This model is able to describe the emergence of cycloidal spin phase in TbMnO$_3$ with a relatively strong GdFeO$_3$-type distortion. However, this model is oversimplified which cannot describe rich magnetoelectric phase diagrams and cross-correlation phenomena of the $R$MnO$_3$ system. The shortages of this model will be argued later.

\subsection{Phase diagrams and multiferroic phases}
\begin{figure*} [tb]
\begin{center}
\includegraphics[scale=1.0]{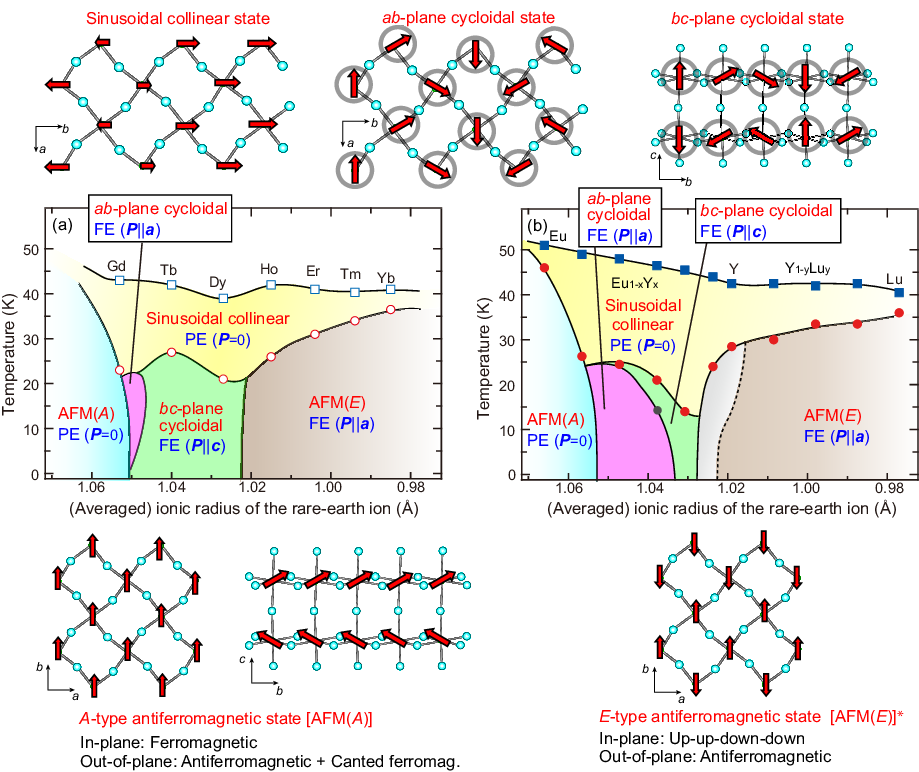}
\caption{(Color online) (a) Experimentally obtained magnetoelectric phase diagrams in plane of (averaged) radius of the $R$ ion and temperature for end compounds $R$MnO$_3$ and solid solutions Gd$_{1-x}$Tb$_x$MnO$_3$. (b) That for solid solutions Eu$_{1-x}$Y$_x$MnO$_3$ and Y$_{1-y}$Lu$_x$MnO$_3$~\cite{Ishiwata10}. Schematic figures of the Mn-spin structures in respective phases are also shown. The $E$-type antiferromagnetic state is described as a collinear spin structure here according to the naive belief in the early stage of the research, but it was theoretically revealed later that the Mn-spins in this phase are canted and form a strongly depressed cycloidal order with fourfold periodicity in reality~\cite{Mochizuki10a}. This figure is taken and modified from Ref.~\cite{Mochizuki11a} {\copyright} 2011 American Physical Society.}
\label{Fig09}
\end{center}
\end{figure*}
An experimental phase diagram of $R$MnO$_3$ is shown in Fig.~\ref{Fig09}(a)~\cite{Ishiwata10}. Here the behaviors between Tb and Gd are investigated using solid solutions Gd$_{1-x}$Tb$_x$MnO$_3$~\cite{Goto05}. When the GdFeO$_3$-type lattice distortion increases with decreasing size of $R$ ion, the $A$-type antiferromagnetic phase, the $ab$-plane cycloidal spin phase, the $bc$-plane cycloidal spin phase, and the $E$-type antiferromagnetic phase appear successively at low temperatures. In addition, a spin-density-wave phase with sinusoidally modulated collinear spins (referred to as sinusoidal collinear phase hereafter) spreads at intermediate temperatures to cover these phases. Among these phases, the $ab$-plane cycloidal phase, the $bc$-plane cycloidal phase, and the $E$-type antiferromagnetic phase are ferroelectric with magnetism-induced ferroelectric polarization, while the other phases are paraelectric. A similar phase diagram was obtained for solid solutions Eu$_{1-x}$Y$_x$MnO$_3$ and Y$_{1-y}$Lu$_y$MnO$_3$ [Fig.~\ref{Fig09}(b)]~\cite{Ishiwata10}. 

Each phase is characterized by the Mn-spin alignment patterns in the $ab$ plane, while the spins are stacked in a staggered manner along the $c$ axis because of the strong inter-plane antiferromagnetic exchange interactions $J_c$. The $A$-type antiferromagnetic phase has ferromagnetically aligned Mn spins in the $ab$ plane, which are oriented approximately along the $b$ axis but are slightly canted along the $c$ axis due to the Dzyaloshinskii-Moriya interaction. The $ab$-plane ($bc$-plane) cycloidal spin phase has a transverse spiral configuration of Mn spins rotating within the $ab$ ($bc$) plane, which propagates along the $b$ axis with incommensurate periodicity. The $E$-type antiferromagnetic phase has an up-up-down-down configuration of Mn spins with fourfold periodicity. The sinusoidal collinear spin phase at intermediate temperatures is a sinusoidally modulated spin-density wave of Mn spins which are collinearly aligned ($\parallel$$\bm b$) with incommensurate periodicity.

In these phase diagrams, a magnetic phase transition with 90-degree flop of the spin cycloidal plane from $ab$ to $bc$ occurs with the decrease of ionic radius of the $R$ ion or with the increase of the GdFeO$_3$-type lattice distortion. The mechanism of this lattice-distortion-induced cycloidal-plane flop (or ferroelectric polarization flop) is not trivial~\cite{Mochizuki09a}. In compounds with small rare-earth ions (i.e., Y, Ho, $\cdots$ , Lu) and significant GdFeO$_3$-type distortion, the $E$-type antiferromagnetic phase with a fourfold periodic Mn-spin order is realized. In this $E$-type antiferromagnetic phase, the emergence of magnetism-induced ferroelectricity was theoretically predicted~\cite{Sergienko06b,Picozzi07,Yamauchi08} and experimentally observed~\cite{Ishiwata10,Lorenz07,Pomjakushin09}. The ferroelectric polarization $\bm P$$\parallel$$\bm a$ in this phase is induced by the uniform oxygen displacement due to the symmetric magnetostriction mechanism. More specifically, the positional shifts of oxygen ions on the Mn-O-Mn bonds along the $x$ ($y$) axis give the electric polarization oriented in the $y$ ($x$) direction, i.e., $\bm P$$\parallel$$\bm y$ ($\bm P$$\parallel$$\bm x$). The ferroelectric polarization $\bm P$$\parallel$$\bm a$ shows up as a sum of these two contributions. The local contribution from the Mn$_i$-O-Mn$_j$ bond to the ferroelectric polarization is proportional to the inner product of the Mn spins $\bm S_i \cdot \bm S_j$. However, the emergence of this $E$-type antiferromagnetic phase itself is not trivial.

\section{Magnetic-Field Control of Ferroelectricity}
\begin{figure} [tb]
\begin{center}
\includegraphics[scale=1.0]{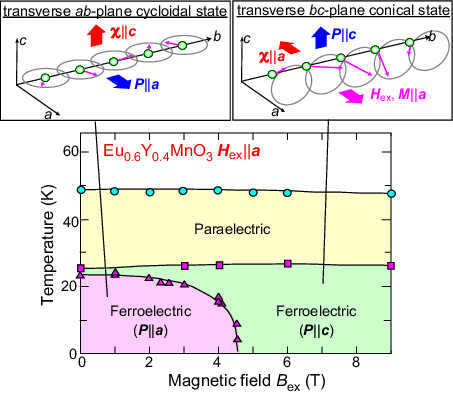}
\caption{(Color online) Experimental phase diagram of Eu$_{0.6}$Y$_{0.4}$MnO$_3$ in a magnetic field $\bm H_{\rm ex}$$\parallel$$\bm a$~\cite{Yamasaki07b}. In the magnetic field, a conical spin order with a uniform spin moment along the field is expected to be stabilized. The observed field-induced 90-degree flop of ferroelectric polarization from $\bm P$$\parallel$$\bm a$ to $\bm P$$\parallel$$\bm c$ can be attributed to a field-induced phase transition from the $ab$-plane cycloidal phase to the transverse $bc$-plane conical phase.}
\label{Fig10}
\end{center}
\end{figure}
In multiferroic materials, the electricity can be controlled by using magnetic field through changing the magnetic structure by taking advantage of the magnetoelectric coupling. When a magnetic field is applied to a cycloidal spin state, a conical spin state with a uniform spin moment along the magnetic field is expected to be stabilized by the energy gain associated with the Zeeman interaction~\cite{Mostovoy06}. In this case, the ferroelectric polarization changes its direction to be perpendicular to the magnetic field upon the reorientation of the cycloidal/conical plane. The experimentally observed 90-degree flop from $\bm P$$\parallel$$\bm a$ to $\bm P$$\parallel$$\bm c$, which occurs when the magnetic field $\bm H_{\rm ex}$$\parallel$$\bm a$ is applied to Eu$_{1-x}$Y$_x$MnO$_3$ in Fig.~\ref{Fig10}, can be understood by this mechanism~\cite{Yamasaki07b}. In other words, it is attributable to a magnetic phase transition from the $ab$-plane cycloidal phase at zero field to the $bc$-plane conical phase in $\bm H_{\rm ex}$$\parallel$$\bm a$. The field-induced reorientation and magnetic-field-based manipulation of ferroelectric polarization with this mechanism have been observed or demonstrated experimentally in TbMnO$_3$~\cite{Abe07}, Eu$_{1-x}$Y$_x$MnO$_3$~\cite{Murakawa08} and many other multiferroic materials, e.g., ZnCr$_2$Se$_4$~\cite{Murakawa08b} and Y-type hexaferrite Ba$_{0.5}$Sr$_{1.5}$Zn$_2$Fe$_{12}$O$_{22}$~\cite{Ishiwata08}. This phenomenon can be reproduced by the simple $J_1$-$J_2$ model through adding the Zeeman-interaction term to the Hamiltonian in Eq.~(\ref{eqn:J1J2M}).

\begin{figure} [tb]
\begin{center}
\includegraphics[scale=1.0]{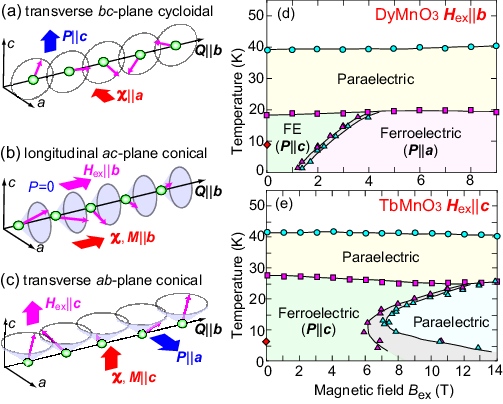}
\caption{(Color online) (a) Transverse $bc$-plane spin cycloid realized in DyMnO$_3$ and TbMnO$_3$ at zero magnetic field. Here $\bm \chi$=$\frac{1}{N}\sum_{\langle i,j \rangle}\bm S_i \times \bm S_j$ is the vector spin chirality, and $\bm Q$($\parallel$$\bm b$) is the propagation vector of spin cycloid. (b) When a magnetic field $\bm H_{\rm ex}$$\parallel$$\bm b$ is applied parallel to $\bm Q$, the longitudinal $ac$-plane conical spin state is naively expected, in which the ferroelectricity should be absent ($\bm P$=0) according to the spin-current model. (c) When $\bm H_{\rm ex}$$\parallel$$\bm a$ is applied perpendicular to $\bm Q$, the transverse $ab$-plane conical spin state with $\bm P$$\parallel$$\bm a$ is naively expected. (d) Experimental phase diagram of DyMnO$_3$ in $\bm H_{\rm ex}$$\parallel$$\bm b$~\cite{Kimura05}. In contrast to the naively expected vanishing ferroelectricity, a 90-degree flop from $\bm P$$\parallel$$\bm c$ to $\bm P$$\parallel$$\bm a$ is observed. (e) Experimental phase diagram of TbMnO$_3$ in $\bm H_{\rm ex}$$\parallel$$\bm c$~\cite{Kimura05}. In contrast to the naively expected $\bm P$$\parallel$$\bm a$ phase, a transition from $\bm P$$\parallel$$\bm c$ to the paraelectric phase with $\bm P$=0 occurs where the ferroelectric polarization vanishes. This figure is taken and modified from Ref.~\cite{Mochizuki10b} {\copyright} 2010 American Physical Society.}
\label{Fig11}
\end{center}
\end{figure}
However, there are some cases that cannot be understood within this simple picture. Here we introduce several strange polarization behaviors observed experimentally in DyMnO$_3$ and TbMnO$_3$~\cite{Kimura05}. DyMnO$_3$ exhibits the $bc$-plane cycloidal spin order propagating along the $b$ axis with $\bm P$$\parallel$$\bm c$ at zero magnetic field [Fig.~\ref{Fig11}(a)]. When a magnetic field $\bm H_{\rm ex}$$\parallel$$\bm b$ is applied, we naively expect that the longitudinal $ac$-plane conical spin order is stabilized [Fig.~\ref{Fig11}(b)]. With this magnetic state, we expect a paraelectric phase with $\bm P$=0 from Eq.~(\ref{eqn:KNB}) because $\bm S_i$$\times$$\bm S_j$ and $\bm e_{ij}$ are parallel. However, a ferroelectric phase with $\bm P$$\parallel$$\bm a$ appears in reality. This means that a nontrivial 90-degree flop from $\bm P$$\parallel$$\bm c$ to $\bm P$$\parallel$$\bm a$ occurs in DyMnO$_3$ under application of $\bm H_{\rm ex}$$\parallel$$\bm b$ [Fig.~\ref{Fig11}(d)].

In fact, such a 90-degree flop of ferroelectric polarization under $\bm H_{\rm ex}$$\parallel$$\bm b$ was also observed in TbMnO$_3$. Similar to DyMnO$_3$, TbMnO$_3$ exhibits the $bc$-plane cycloidal spin order with $\bm P$$\parallel$$\bm c$ at zero field, and the 90-degree flop to $\bm P$$\parallel$$\bm a$ occurs when $\bm H_{\rm ex}$$\parallel$$\bm b$ is applied. A neutron-scattering experiment revealed that the magnetic order in this field-induced $\bm P$$\parallel$$\bm a$ phase is the $ab$-plane spiral~\cite{Aliouane09}. In fact, TbMnO$_3$ exhibits another interesting behavior. When a magnetic field $\bm H_{\rm ex}$$\parallel$$\bm c$ is applied to TbMnO$_3$, we naively expect the transverse $ab$-plane conical spin state with $\bm P$$\parallel$$\bm a$ [Fig.~\ref{Fig11}(c)]. In reality, however, the ferroelectric polarization disappears, and the system becomes paraelectric [Fig.~\ref{Fig11}(e)].

In contrast to the trivial behavior in Fig.~\ref{Fig10} for Eu$_{0.6}$Y$_{0.4}$MnO$_3$, the nontrivial behaviors in Figs.~\ref{Fig11}(d) and (e) observed in DyMnO$_3$ and TbMnO$_3$ cannot be reproduced by the simple $J_1$-$J_2$ model in Eq.~(\ref{eqn:J1J2M}). Because keen competitions among the spin-exchange interactions, the Dzyaloshinskii-Moriya interactions, and magnetic anisotropies under an external magnetic field are relevant to these nontrivial behaviors, theoretical descriptions based on a more elaborate microscopic model are substantially required. Similar nontrivial field-induced behaviors of ferroelectricity have been observed not only for $R$MnO$_3$ but also for many other spin-spiral-based multiferroic materials such as LiCu$_2$O$_2$~\cite{SPark07}, TbMn$_2$O$_5$~\cite{Hur04}, Ni$_3$V$_2$O$_8$~\cite{Kenzelmann06}, and MnWO$_4$~\cite{Taniguchi08}.

\section{Microscopic Model}
The simplest model that describes the frustration in the $ab$-plane of $R$MnO$_3$ is the $J_1$-$J_2$ classical Heisenberg model on the square lattice. This model can reproduce both the two-dimensional ferromagnetic order in the $A$-type antiferromagnetic phase and the cycloidal order. Moreover, the lattice distorion-induced phase transition between these two phases observed in the experimental phase diagrams in Figs.~\ref{Fig09}(a) and (b) can be reproduced by this simple model upon the variation of $J_2$. However, the $R$MnO$_3$ system shows many strange but interesting phenomena which cannot be reproduced by this simple model, which include (1) the 90-degree flop of cycloidal plane or ferroelectric polarization induced by the GdFeO$_3$-type lattice distortion [Figs.~\ref{Fig09}(a) and (b)]~\cite{Mochizuki09a,Mochizuki09b}, (2) the emergence of the multiferroic $E$-type antiferromagnetic phase with $\bm P$$\parallel$$\bm a$ in the strongly distorted materials with $R$=Ho, $\cdots$, Lu~\cite{Mochizuki10a,Mochizuki11a}, (3) the magnetic-field-induced flop and disappearance of ferroelectric polarization in DyMnO$_3$ and TbMnO$_3$ [Figs.~\ref{Fig11}(d) and (e)]~\cite{Mochizuki10b,Matsubara15,Mochizuki15}, and (4) the electromagnon excitations activated by light electric field~\cite{Mochizuki10c}.

These phenomena indicate that necessity to construct a more elaborate model. In the frustrated system, the energy scale of the spin-exchange interactions becomes effectively small, so that weak magnetic anisotropies and magnetic interactions originating from the spin-orbit coupling and the spin-lattice coupling get relative importance to play significant roles in the physical phenomena. To describe the magnetoelectric phenomena in $R$MnO$_3$, we construct a model in which such situations are properly taken into account. First, to reproduce the 90-degree flop of the spin cycloidal plane from $ab$ to $bc$, we consider competing magnetic interactions and magnetic anisotropies relevant to stability of the respective spin cycloids. Here we consider the Dzyaloshinskii-Moriya interactions with vectors on the Mn-O-Mn bonds with broken spatial inversion symmetry due to the GdFeO$_3$-type lattice distortion. We also consider the site-dependent single-ion magnetic anisotropies reflecting the staggered orbital ordering and tilting of the MnO$_6$ octahedra. In addition, we expect that the $E$-type (up-up-down-down type) antiferromagnetic state with both antiferromagnetically and ferromagnetically coupled spins strongly couples with the bond-alternation type lattice distortion. Therefore, we consider the Peierls-type spin-lattice coupling and the lattice elastic energy. Specifically, we construct the following model~\cite{Mochizuki09a,Mochizuki09b,Mochizuki10a,Mochizuki11a}.
\begin{equation}
\mathcal{H}=
\mathcal{H}_{\rm ex}+\mathcal{H}_{\rm sia}^D+\mathcal{H}_{\rm sia}^E+\mathcal{H}_{\rm DM}+\mathcal{H}_K,
\label{eqn:Hamlt}
\end{equation}
with
\begin{eqnarray}
\mathcal{H}_{\rm ex}&=&
\sum_{<i,j>} J_{ij} \bm S_i \cdot \bm S_j,\\
\label{eqn:HamltJ}
\mathcal{H}_{\rm sia}^D&=&D\sum_{i}S_{\zeta i}^2,\\
\label{eqn:HamltD}
\mathcal{H}_{\rm sia}^E&=&
E\sum_{i}(-1)^{i_x+i_y}(S_{\xi i}^2-S_{\eta i}^2),\\
\label{eqn:HamltE}
\mathcal{H}_{\rm DM}&=&
\sum_{<i,j>}\bm d_{i,j}\cdot(\bm S_i \times \bm S_j),\\
\label{eqn:HamltD}
\mathcal{H}_K&=&K \sum_i(\delta_{i,i+\hat x}^2+\delta_{i,i+\hat y}^2),
\label{eqn:HamltK}
\end{eqnarray}

In this model, we consider classical spin vectors ${\bm S}_i$ on a orthorhombic lattice,
\begin{eqnarray}
{\bm S}_i=(\sqrt{S^2-S_{ci}^2}\cos\theta_i, \sqrt{S^2-S_{ci}^2}\sin\theta_i, S_{ci}) .
\end{eqnarray}
The magnitude of $S$ is taken to be $S$=2 to reproduce the magnetic moment 4$\mu_{\rm B}$ of the Mn$^{3+}$ ion. Here ($i_x$, $i_y$, $i_z$) are integer coordinates of the $i$th Mn site. We consider the following exchange interactions $J_{ij}$ between Mn spins: (1) $J_{ab}(<0)$ for the in-plane nearest-neighbor ferromagnetic exchange interactions along the $x$ and $y$ axes, (2) $J_b(>0)$ for the in-plane next-neighbor antiferromagnetic exchange interactions along the $b$ axis, and (3) $J_c(>0)$ for the inter-plane antiferromagnetic exchange interactions along the $c$ axis. A very weak ferromagnetic interactions $J_a<0$ are also considered for the Mn-Mn bonds along the $a$ axis to reproduce the $E$-type antiferromagnetic phase.

In addition to these exchange interactions ($\mathcal{H}_{\rm ex}$), we consider the single-ion magnetic anisotropy terms ($\mathcal{H}_{\rm sia}^D$ and $\mathcal{H}_{\rm sia}^E$) and the Dzyaloshinskii-Moriya interactions ($\mathcal{H}_{\rm DM}$). They all come from the relativistic spin-orbit coupling. Here single-ion anisotropies are defined by adopting the local coordinate axes $\xi_i$, $\eta_i$, and $\zeta_i$ attached to each MnO$_6$ octahedron to reflect the structure and nature of the 3$d$ orbitals on each Mn ion governed by the local octahedral crystal fields [see Fig.~\ref{Fig07}(a)]. The directional vectors $\bm \xi_i$, $\bm \eta_i$ and $\bm \zeta_i$ with respect to the $a$, $b$ and $c$ axes are given by
\begin{eqnarray}
\label{eq:tilax1}
\bm {\xi}_i&=&\left[
\begin{array}{c}
a[0.25+(-1)^{i_x+i_y}(0.75-x_{{\rm O}_2})] \\
b[0.25-(-1)^{i_x+i_y}(y_{{\rm O}_2}-0.25)] \\
c(-1)^{i_x+i_y+i_z}z_{{\rm O}_2} \\
\end{array}
\right], \\
\label{eq:tilax2}
\bm {\eta}_i&=&\left[
\begin{array}{c}
a[-0.25+(-1)^{i_x+i_y}(0.75-x_{{\rm O}_2})] \\
b[0.25+(-1)^{i_x+i_y}(y_{{\rm O}_2}-0.25)] \\
-c(-1)^{i_x+i_y+i_z}z_{{\rm O}_2} \\
\end{array}
\right], \\
\label{eq:tilax3}
\bm {\zeta}_i&=&\left[
\begin{array}{c}
-a(-1)^{i_x+i_y+i_z}x_{{\rm O}_1} \\
b(-1)^{i_z}(0.5-y_{{\rm O}_1}) \\
0.25c \\
\end{array}
\right].
\end{eqnarray}
Here $x_{{\rm O}_2}$, $y_{{\rm O}_2}$ and $z_{{\rm O}_2}$ ($x_{{\rm O}_1}$ and $y_{{\rm O}_1}$) are the coordination parameters of the in-plane (inter-plane) oxygen ions, and $a$, $b$ and $c$ are the lattice parameters. Consideration of the site-specific tilted coordinate axes is essential not only to improve the quantitative accuracy of the calculations but also to reproduce the emergence of sinusoidal collinear phase at intermediate temperatures and the easy-magnetization axis ($\parallel$$\bm b$) in the $A$-type antiferromagnetic and collinear magnetic phases. These local coordinate axes are calculated using the experimentally determined structural parameters~\cite{Mochizuki09a}.

\begin{figure} [tb]
\begin{center}
\includegraphics[scale=1.0]{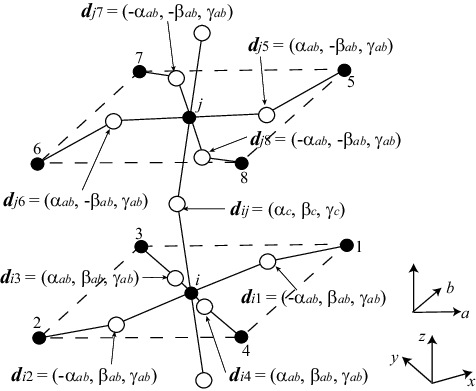}
\caption{Dzyaloshinskii-Moriya vectors on the Mn-O-Mn bonds. The vectors on the in-plane Mn-O-Mn bonds are represented by three parameters, $\alpha_{ab}$, $\beta_{ab}$, $\gamma_{ab}$, while the vectors on the inter-plane Mn-O-Mn bonds are represented by two parameters, $\alpha_c$ and $\beta_c$.}
\label{Fig12}
\end{center}
\end{figure}
The Dzyaloshinskii-Moriya vectors on the Mn-O-Mn bonds are oriented nearly pependicular to the local Mn-O-Mn planes, and are described by five parameters because of the crystal symmetry~\cite{Solovyev96}. The vectors on the in-plane bonds are described by three parameters $\alpha_{ab}$, $\beta_{ab}$, and $\gamma_{ab}$. In contrast, the vectors on the inter-plane bonds are described by two parameters $\alpha_c$ and $\beta_c$, where the $c$-axis component is always zero because of the mirror symmetry. Their expressions are given by [see also Fig.~\ref{Fig12}],
\begin{eqnarray}
\bm d_{i,i+\hat x}&=&\left[
\begin{array}{c}
-(-1)^{i_x+i_y+i_z}\alpha_{ab} \\
(-1)^{i_x+i_y+i_z}\beta_{ab} \\
(-1)^{i_x+i_y}\gamma_{ab} \\
\end{array}
\right], \\
\bm d_{i,i+\hat y}&=&\left[
\begin{array}{c}
(-1)^{i_x+i_y+i_z}\alpha_{ab} \\
(-1)^{i_x+i_y+i_z}\beta_{ab} \\
(-1)^{i_x+i_y}\gamma_{ab} \\
\end{array}
\right], \\
\bm d_{i,i+\hat z}&=&\left[
\begin{array}{c}
(-1)^{i_z}\alpha_c \\
(-1)^{i_x+i_y+i_z}\beta_c \\
0 \\
\end{array}
\right].
\label{eq:DMVECS}
\end{eqnarray}

Furthermore, because the nearest-neighbor ferromagnetic exchange interactions $J_{i,j}$ in $R$MnO$_3$ sensitively depend on the Mn$_i$-O-Mn$_j$ bond angle, we consider the following Peierls-type spin-lattice coupling,
\begin{align}
J_{i,j}=J_{ab}+J'_{ab}\delta_{i,j},
\label{eqn:Peierls}
\end{align}
where $J_{ab}'$=$\partial J_{ab}$/$\partial \delta$. The $\delta_{i,j}$ and $\delta$ are the displacement of oxygen along the $\bm n_{i,j}$ axis connecting the midpoint of two Mn ions (Mn$_i$ and Mn$_j$) and the oxygen, which is normalized by the averaged MnO bond length. The last term $\mathcal{H}_{\rm elas}$ is the elastic energy term associated with oxygen displacements where $K$ is the elastic constant.

\begin{table}[tbp]
\caption{Model parameters~\cite{Mochizuki10a} and crystal parameters~\cite{Alonso00} used in the calculations for the magnetoelectric phase diagram in Fig.~\ref{Fig13}. The energy unit of the model parameters is meV.}
\begin{tabular}{c|llll}
\hline
\hline
$\mathcal{H}_{\rm ex}$ & $J_{ab}=-0.8$, & $J_a=-0.1$, & $J_c$=1.25 & \\
                       & $J_{ab}'=2$    &             &            & \\
$\mathcal{H}_{\rm sia}^D$, $\mathcal{H}_{\rm sia}^E$ &
$D$=0.2,      & $E$=0.25   & & \\
$\mathcal{H}_{\rm DM}$  &
$\alpha_{ab}$=0.1, & $\beta_{ab}$=0.1, & $\gamma_{ab}$=0.14 \\
                        &
$\alpha_c$=0.42, & $\beta_c$=0.1 &  \\
$\mathcal{H}_{\rm elas}$ & $K$=500    &            & & \\
\hline
 & $a$=5.2785 $\AA$ & $b$=5.2785 $\AA$ & $c$=7.3778 $\AA$  & \\
 & $x_{\rm O_1}$=0.1092 & $y_{\rm O_1}$=0.4642 &  & \\
 & $x_{\rm O_2}$=0.7028 & $y_{\rm O_2}$=0.3276 &  $z_{\rm O_2}$=0.0521 & \\
\hline
\hline
\end{tabular}
\label{tabl:MDLPRMS}
\end{table}
\begin{figure} [tb]
\begin{center}
\includegraphics[scale=1.0]{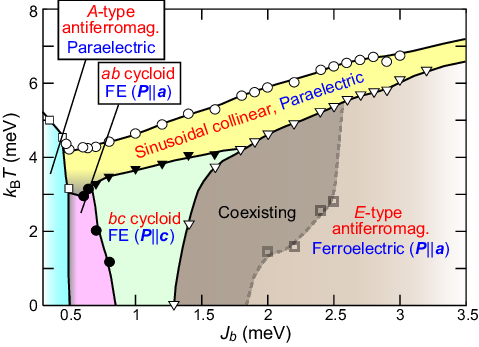}
\caption{(Color online) Theoretical magnetoelectric phase diagram of the spin model in Eq.~(\ref{eqn:Hamlt}) in the plane of $J_b$ and temperature obtained by the Monte Carlo calculations. This figure is taken and modified from Ref.~\cite{Mochizuki11a} {\copyright} 2011 American Physical Society.}
\label{Fig13}
\end{center}
\end{figure}
This spin-lattice coupled model was analyzed using the replica exchange Monte Carlo method to investigate the magnetoelectric phase diagram of $R$MnO$_3$ in Fig.~\ref{Fig13}. Among the model parameters, $J_{ab}$, $J_{ab}'$, $J_b$, $J_c$, $D$, and $E$ were determined by microscopic calculations~\cite{Mochizuki09a,Mochizuki10a}. The five Dzyaloshinskii-Moriya parameters were determined so as to reproduce the phase diagrams at zero magnetic field and those in magnetic fields based on results of the first-principles band calculations~\cite{Solovyev96} and the electron spin resonance experiments~\cite{Tovar99,Deisenhofer02}. The value of the elastic constant $K$ was determined to reproduce the experimental value of the electric polarization originating from the oxygen displacement in the $E$-type antiferromagnetic phase. For more details of the parameter evaluations, see Refs.~\cite{Mochizuki09a,Mochizuki10a}. All of these parameters except $J_b$ are almost constant in the region of moderate lattice distortion where the multiferroic phases with cycloidal spin orders take place. In contrast, the value of $J_b$ varies sensitively depending on the degree of GdFeO$_3$-type lattice distortion. Therefore, we fixed the parameters other than $J_b$ at the constant values listed in Table~\ref{tabl:MDLPRMS} and treated only $J_b$ as a variable. The local coordinate axes were calculated using the experimental crystal parameters for DyMnO$_3$~\cite{Alonso00} listed in Table~\ref{tabl:MDLPRMS} throughout the calculations.

Figure~\ref{Fig13} shows calculated magnetoelectric phase diagram in the plane of $J_b$ and temperature. This theoretical phase diagram reproduces well the experimental phase diagrams shown in Figs.~\ref{Fig09}(a) and (b). Note that the phase diagram in Fig.~\ref{Fig09}(b) for the solid-solution systems shows that the transition temperature is strongly suppressed at the phase boundary between the $bc$-plane cycloidal phase and the $E$-type antiferromagnetic phase. This suppression of transition temperature at the phase boundary is attributable to the disorder effect inherent in the solid solution systems~\cite{Tomioka04}. Conversely, only a tiny dip appears at the phase boundary in the phase diagram of the end compounds $R$MnO$_3$ in Fig.~\ref{Fig09}(a) where the disorder effect is almost absent. In the phase diagram in Fig.~\ref{Fig13} obtained by the theoretical calculations without considering disorders, an almost straight phase boundary with no anomaly is obtained. 

It was revealed that the spin configuration in the $E$-type antiferromagnetic phase is not a naively believed simple collinear up-up-down-down type but a significantly depressed $ab$-plane cycloid with a large ellipticity. It is also worth mentioning that there exists an area in which the (incommensurate) $ab$-plane cycloidal spin phase and the commensurate $E$-type antiferromagnetic phase coexist near the phase boundary between these phases. This coexistence can explain the contradicting results of neutron-scattering experiments about the magnetic wavenumbers in YMnO$_3$ and HoMnO$_3$, that is, some experiments indicated incommensurate wavenumbers, while the other experiments indicated a commensurate wavenumber of $q_b=0.5\pi$ corresponding to the four-fold periodicity~\cite{Munoz01,Brinks01,Munoz02,YeF07}. 

\section{Dynamical Magnetoelectric Phenomena}
The magnetoelectric coupling in multiferroics also gives rise to interesting cross-correlation phenomena in the dynamical regime. We discuss the electromagnon excitations in the multiferroic Mn perovskites and the theoretically proposed optical switching of spin chirality utilizing their intense excitations.

\subsection{Electromagnon excitations}
The possibility of spin-wave excitations in multiferroics through exciting the electric polarizations with light electric field has been proposed both theoretically~\cite{Smolenski82,Katsura07} and experimentally~\cite{Pimenov06a,Pimenov06b}. In addition to the scientific importance as novel magnon excitations coupled with electric dipoles, the electromagnons are of interest from the viewpoint of potential applications such as optical control of magnetoelectricity and optical device functions. Furthermore, the strong coupling between electric polarization and electric fields, which is several orders of magnitude stronger than that between magnetization and magnetic fields, enables intense excitation of (electro)magnons that cannot be realized for conventional magnons excited by magnetic fields. Therefore, dramatic nonlinear dynamics, cooperative phenomena, and switching phenomena can be expected for the electromagnon excitations. It might open a novel magnon physics in the nonlinear regime associated with unique phenomena such as magnon Bose-Einstein condensations, photoinduced magnetic phase transitions, magnon lasers, and magnon Cherenkov radiations as research targets.

\begin{figure} [tb]
\begin{center}
\includegraphics[scale=1.0]{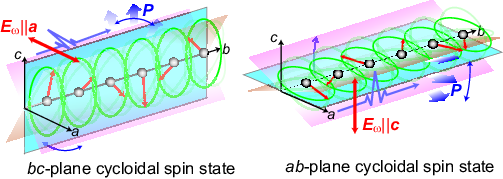}
\caption{(Color online) Physical mechanism of the electromagnon excitations in $R$MnO$_3$ proposed in the early stage of the research. The light electric field induces oscillations of the ferroelectric polarizations and causes linked oscillation of the spin spiral plane. For this mechanism, the collective mode can be excited only when the light electric field is perpendicular to the spin cycloidal plane, which is not compatible with the experimental observation. Specifically, the electromagnon excitation is expected with $\bm E_\omega$$\parallel$$\bm a$ for the $bc$-plane cycloidal phase with $\bm P$$\parallel$$\bm c$ (left panel), while it is expected with $\bm E_\omega$$\parallel$$\bm c$ for the $ab$-plane cycloidal phase with $\bm P$$\parallel$$\bm a$ (right panel).}
\label{Fig14}
\end{center}
\end{figure}
\begin{figure} [tb]
\begin{center}
\includegraphics[scale=1.0]{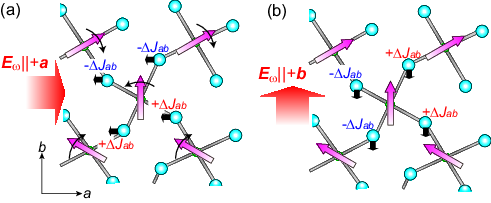}
\caption{(Color online) (a) Newly proposed physical mechanism of the electromagnon excitations~\cite{Mochizuki10c}. The local electric polarizations caused by displaced oxygen ions on the zigzag Mn-O-Mn bonds couple with the light electric field to modulate the nearest-neighbor ferromagnetic exchange interactions. In this case, only the modulation realized by the light electric field parallel to the $a$ axis can activate the collective spin excitation, which is consistent with the experimental observation. (b) Modulation of the nearest-neighbor ferromagnetic exchange interactions when the light electric field is parallel to the $b$ axis, which cannot activate the collective spin excitation.}
\label{Fig15}
\end{center}
\end{figure}
An excitation mechanism of the electromagnons in $R$MnO$_3$ proposed in the early stage of the research was that the light electric field excites the ferroelectric polarization induced by the cycloidal Mn-spin order to cause oscillations of the cycloidal spin plane around the propagation vector (parallel to the $b$ axis for the present case) as shown in Fig.~\ref{Fig14}~\cite{Katsura07}. In this case, there should be a selection rule whether the light electric field can excite this rotational mode, which depend on the relative orientations of the cycloidal spin plane and the light electric field. Specifically, when the cycloidal plane is $bc$ as shown in the left panel of Fig.~\ref{Fig14}, the light electric field parallel to the $a$ axis (i.e., $\bm E_\omega$$\parallel$$\bm a$) can shake the ferroelectric polarization $\bm P$$\parallel$$\bm c$ to excite the rotational oscillation of the cycloidal plane. In contrast, when the cycloidal plane is $ab$ as shown in the right panel of Fig.~\ref{Fig14}, the light electric field must be parallel to the $c$ axis (i.e., $\bm E_\omega$$\parallel$$\bm c$) to excite the rotational oscillation of the cycloidal plane. Namely, such a magnon mode can be expected only when the light electric field is perpendicular to the cycloidal spin plane. In the optical experiments for $R$MnO$_3$, however, the optical absorptions associated with electromagnon excitations are observed only when the light electric field was parallel to the $a$ axis, irrespective of the orientation of the cycloidal plane~\cite{KidaRV09,Kida08b}. These observations apparently contradict to the above naive picture.

Subsequently, a new excitation mechanism was proposed~\cite{Mochizuki10a,Aguilar09}. The mechanism is that a light electric field excites local electric polarizations existing in the bended Mn-O-Mn bonds on the zigzag chains. These local electric polarizations cancel out in total because of their staggered configurations and do not contribute to the ferroelectric polarization. However, their oscillatory dynamics induced by light electric field can give rise to electromagnon excitations. As shown in Figs.~\ref{Fig15}(a) and (b), the O$^{2-}$ ions are displaced from their original positions by the light electric field in the directions indicated by short thick arrows. The oxygen displacements increase or decrease the Mn-O-Mn bond angles. In $R$MnO$_3$, the strength of the local nearest-neighbor ferromagnetic exchange interaction $J_{ab}(<0)$ depends sensitively on the bond angle. Specifically, the ferromagnetic interaction becomes stronger as $J_{ab} \rightarrow J_{ab}+\Delta J_{ab}$ ($\Delta J_{ab}<0$) when the Mn-O-Mn bond angle increases and approaches 180$^\circ$, while it becomes weaker as $J_{ab} \rightarrow J_{ab}-\Delta J_{ab}$ when the bond angle decreases. 

Figures ~\ref{Fig15}(a) and (b) show the way of the modulations of local ferromagnetic exchange interactions induced by the electric fields $\bm E_\omega$$\parallel$$+\bm a$ and $\bm E_\omega$$\parallel$$+\bm b$, respectively. Note that the sign reversal of the light electric field reverses the sign of the modulations. According to these figures, we find that the way of the modulation depends on the direction of  $\bm E_\omega$ or the light polarization. Note that because the angle between two spins connected by a bond with enhanced (weakened) ferromagnetic interaction should become smaller (larger), the application of electric field causes rotations of the spins. Only modulation of the interactions under $\bm E_\omega$$\parallel$$\bm a$ can induce cooperative spin rotations as indicated by thin arrows in Fig.~\ref{Fig15}(a), whereas that under $\bm E_\omega$$\parallel$$\bm b$ cannot. This is consistent with the experimental observation that the electromagnon excitations show up only when the light polarization is $\bm E_\omega$$\parallel$$\bm a$.

To verify this mechanism, the optical absorption spectra and magnon dispersion relations were calculated by incorporating the microscopic model of $R$MnO$_3$ into the time evolution equation of spins~\cite{Mochizuki10a}. More specifically, spatiotemporal profiles of the Mn-spin dynamics after application of the light electric-field pulse were simulated by numerically solving the Landau-Lifshitz-Gilbert (LLG) equation using the Runge-Kutta method. The LLG equation is given by,
\begin{align}
\frac{\partial \bm S_i}{\partial t}=-\bm S_i \times \bm H^{\rm eff}_i
+ \frac{\alpha_{\rm G}}{S} \bm S_i \times \frac{\partial \bm S_i}{\partial t}.
\label{eq:LLGEQ}
\end{align} 
The first term of this equation is referred to as the gyrotropic term and describes the rotational motion of spins around the effective local magnetic field $\bm H^{\rm eff}_i$. The second term is phenomenologically introduced to describe the energy dissipation and is referred to as the Gilbert-damping term where $\alpha_{\rm G}$ is called the Gilbert-damping coefficient. The effective local magnetic field $\bm H^{\rm eff}_i$ acting on the $i$th spin $\bm S_i$ is calculated by a spin-derivative of the Hamiltonian as,
\begin{align}
\bm H^{\rm eff}_i = - \frac{\partial \mathcal{H}}{\partial \bm S_i}.
\label{eq:EFFMF}
\end{align} 
For the Hamiltonian $\mathcal{H}$, we slightly modified the original Hamiltonian in Eq.~(\ref{eqn:Hamlt}). Specifically, the Peierls-type spin-lattice coupling was not considered by setting the nearest-neighbor ferromagnetic exchange interaction to be constant as $J_{i,j}=J_{ab}$, and the lattice elastic term $\mathcal{H}_K$ was removed. Instead, the interaction term originating from the dynamical spin-phonon coupling called biquadratic term is taken into account~\cite{Mochizuki10c}. The model parameters and the crystal parameters used in the numerical calculations are summarized in Table~\ref{tabl:PRMVLS}. For the model parameters, values evaluated by microscopic calculations were used~\cite{Mochizuki09a}. On the other hand, values obtained by neutron diffraction experiments were used for the crystal parameters~\cite{Alonso00,Dabrowski05}. These parameters accurately reproduce the cycloidal spin states in TbMnO$_3$, and DyMnO$_3$ and Eu$_{1-x}$Y$_x$MnO$_3$ with respect to the orientation of spin cycloidal plane and the magnetic wavenumber. The effect of light-pulse application is treated by short-time modulation of the nearest-neighbor ferromagnetic exchange interactions as shown in Figs. ~\ref{Fig15}(a) and (b).

\begin{table}
\caption{Model parameters~\cite{Mochizuki10a} (in meV) and crystal parameters~\cite{Alonso00,Dabrowski05} for TbMnO$_3$, DyMnO$_3$ and Eu$_{0.5}$Y$_{0.5}$MnO$_3$ used in the calculations for electromagnon excitations. The crystal parameters for Eu$_{0.5}$Y$_{0.5}$MnO$_3$ are evaluated by interpolation  between EuMnO$_3$~\cite{Dabrowski05} and YMnO$_3$~\cite{Alonso00}. Cycloidal planes and magnetic modulation wavenumbers $\bm Q $=(0, $q_b$, $\pi$) of cycloidal spin states produced by these parameter sets are also shown. For each material, the experimentally observed cycloidal plane and magnetic wavenumber can be well reproduced using the parameter set.}
\begin{center}
\begin{tabular}{c|ccc}
\hline
\hline
 $R$MnO$_3$ & TbMnO$_3$ & DyMnO$_3$ & Eu$_{0.5}$Y$_{0.4}$MnO$_3$\\
\hline
$J_{ab}$ &-0.74 &-0.70 &-0.74 \\
$J_b$    & 0.64 & 0.99 & 0.74 \\
$J_c$    & 1.00 & 1.00 & 1.20 \\
$D$      & 0.20 & 0.22 & 0.24 \\
$E$      & 0.25 & 0.25 & 0.30 \\
$\alpha_{ab}$    & 0.10  & 0.10 &0.10 \\
$\beta_{ab}$      & 0.10  & 0.10 &0.10 \\
$\gamma_{ab}$  & 0.14  & 0.14 &0.16 \\
$\alpha_c$    & 0.48  & 0.45 &0.40 \\
$\beta_c$      & 0.10  & 0.10 &0.10 \\
$\gamma_c$  & 0  & 0 &0 \\
$B_{\rm biq}$      & 0.025 & 0.025 & 0.025 \\
\hline
$a$ ($\AA$)      & 5.29314& 5.2785& 5.28256 \\
$b$ ($\AA$)      & 5.8384 & 5.8337 & 5.81618 \\
$c$ ($\AA$)      & 7.4025 & 7.3778 &  7.403324\\
$x_{\rm O_1}$ & 0.1038 & 0.1092 &  0.10454 \\
$y_{\rm O_1}$ & 0.4667 & 0.4642 &  0.46816\\
$x_{\rm O_2}$ & 0.7039 & 0.7028 &  0.70250\\
$y_{\rm O_2}$ & 0.3262 & 0.3276 &  0.32584\\
$z_{\rm O_2}$ & 0.0510 & 0.0521 &  0.05058\\
\hline
Cycloidal plane & $bc$ & $bc$ & $ab$ \\
$q_b$ & 0.3$\pi$  & 0.4$\pi$  & $\pi$/3 \\
$q_b$(Experiment) & 0.28$\pi$ & 0.39$\pi$ & $\sim$$\pi$/3 \\
\hline
\hline
\end{tabular}
\end{center}
\label{tabl:PRMVLS}
\end{table}
Here we explain how to calculate the optical absorption spectrum from the simulated time-profile data of the spin dynamics. Spin oscillations excited by the light electric-field pulse induce oscillating local electric polarizations through the exchange magnetostriction mechanism. When the Mn-O-Mn bond angle and, hence, the nearest-neighbor ferromagnetic exchange interactions are modulated by the oxygen displacements, the angle between two Mn spins connected by the bond changes. As an inverse effect of this phenomenon, when the spin-spin angle changes due to the light-induced spin oscillation, the oxygen between these two spins spontaneously shifts to increase (decrease) the ferromagnetic exchange interaction so as to maximize the energy gain when the spin-spin angle becomes smaller (larger). As a result, increase or decrease of the local electric polarization, which scales with $\bm S_i \cdot \bm S_j$, occurs on the Mn$_i$-O-Mn$_j$ bond. Eventually, a temporally varying net component of the electric polarizations appears, which is give by,
\begin{align}
\bm P_{\rm s}(t)=\sum_{ij} \bm \Pi_{i,j}[\bm S_i(t) \cdot \bm S_j(t)],
\label{eq:PSS}
\end{align}
where $\bm \Pi_{i,j}$ is a form factor for the zigzag Mn-O chains. This mechanism is called the symmetric exchange striction mechanism. From the simulated time profile of spin dynamics, we calculate the time evolution of $\bm P_{\rm s}(t)$ and subsequently calculate the $\omega$-dependence of the dielectric constant $\varepsilon$ by Fourier transformation of $\bm P_{\rm s}(t)$. Its imaginary part, Im$\varepsilon(\omega)$, corresponds to the optical absorption spectrum.

\begin{figure*} [tb]
\begin{center}
\includegraphics[scale=1.0]{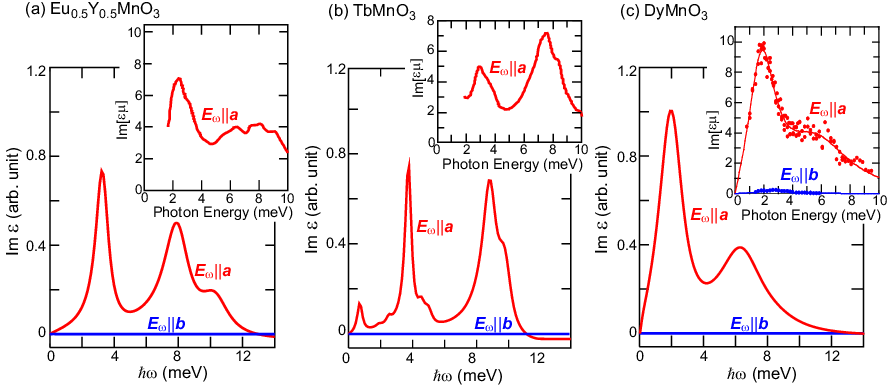}
\caption{(Color online) Calculated optical absorption spectra associated with electromagnon excitations for (a) Eu$_{0.5}$Y$_{0.5}$MnO$_3$ with $ab$-plane cycloidal spin order, (b) TbMnO$_3$ with $bc$-plane cycloidal spin order, and (c) DyMnO$_3$ with $bc$-plane cycloidal spin order. Both the cases with $\bm E_\omega$$\parallel$$\bm a$ and $\bm E_\omega$$\parallel$$\bm b$ are investigated. Two characteristic peaks are observed at terahertz frequencies only when $\bm E_\omega$$\parallel$$\bm a$. Insets show the experimental spectra for Eu$_{0.55}$Y$_{0.45}$MnO$_3$~\cite{Takahashi09}, TbMnO$_3$~\cite{Takahashi08}, and DyMnO$_3$~\cite{Kida08}. Figures (a) and (c) are taken and modified from Ref.~\cite{Mochizuki10c} {\copyright} 2010 American Physical Society.}
\label{Fig16}
\end{center}
\end{figure*}
\begin{figure} [tb]
\begin{center}
\includegraphics[scale=1.0]{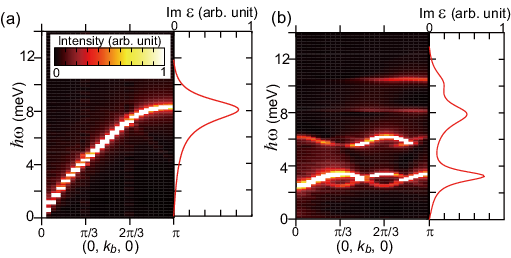}
\caption{(Color online) (a) Calculated spin-wave dispersion relation (left panel) and optical absorption spectrum (right panel) for a uniformly rotating cycloidal Mn-spin order with an equivalent pitch angle produced by a Hamiltonian composed of only the exchange-interaction term $\mathcal{H}=\mathcal{H}_{\rm ex}$. (a) Those for a cycloidal Mn-spin order with nonuniform pitch angles and higher-harmonic components produced by the full Hamiltonian in Eq.~(\ref{eqn:Hamlt}). This figure is taken and modified from Ref.~\cite{Mochizuki10c} {\copyright} 2010 American Physical Society.}
\label{Fig17}
\end{center}
\end{figure}
Figures~\ref{Fig16}(a)-(c) show the calculated optical absorption spectra for (a) Eu$_{1-x}$Y$_x$MnO$_3$ with $ab$-plane cycloidal spin order, (b) TbMnO$_3$ with $bc$-plane cycloidal spin order, and (c) DyMnO$_3$ with $bc$-plane cycloidal spin order. Both light polarizations of $\bm E_\omega$$\parallel$$\bm a$ and $\bm E_\omega$$\parallel$$\bm b$ are investigated. We find that two characteristic peaks are observed in the THz-frequency regime only when the light electric field is parallel to the $a$ axis, i.e., $\bm E_\omega$$\parallel$$\bm a$, irrespective of the orientation of the cycloidal spin plane. These calculated spectra reproduce well the experimental spectra shown in the insets of these figures~\cite{Takahashi09,Takahashi08,Kida08}. Of the two peaks seen in each spectrum for $\bm E_\omega$$\parallel$$\bm a$, the higher-frequency peak at $\omega$$\sim$2 THz ($\sim$8 meV) is attributed to the spin-wave mode indicated by thin arrows in Fig.~\ref{Fig15}(a), which is a zone-edge mode corresponding to the wavenumber ($k_a$, $k_b$, $k_c$)=(0, $\pi$, 0). On the contrary, another peak around $\omega$$\sim$0.7 THz ($\sim$3-4 meV) is attributed to the higher-harmonic mode. 

The cycloidally ordered Mn spins in $R$MnO$_3$ are not uniformly rotating but contains large higher-harmonic components due to the site-dependent magnetic anisotropies and the staggered Dzyaloshinskii-Moriya interactions. As a result, multiple folding of the spin-wave dispersions occurs at the corresponding wavenumbers, which gives rise to peaks at the low-energy regime. In fact, a uniformly rotating spiral spin order produced by a Hamiltonian containing only the spin exchange interactions yields a single spin-wave dispersion [left panel of Fig.~\ref{Fig17}(a)] and an optical absorption spectrum with a single peak at the energy corresponding to the zone-edge mode [right panel of Fig.~\ref{Fig17}(a)]. In contrast, the nonuniformly rotating cycloidal spin order with higher-harmonic components produced by the full Hamiltonian in Eq.~(\ref{eqn:Hamlt}), which incorporates all the interactions and anisotropies, yields multiple foldings and complicated anticrossings in the spin-wave dispersions [left panel of Fig.~\ref{Fig17}(b)]. With these spin-wave dispersions, a low-energy peak appears in addition to the high-energy peak [right panel of Fig.~\ref{Fig17}(b)].

\section{Proposed Optical Switching of Spin Chirality}
Our next interest turns to nonlinear dynamical phenomena associated with intense electromagnon excitations~\cite{Kubacka2014}. We here discuss a theoretically predicted possible optical switching of spin chirality by electromagnon excitations with intense laser irradiation~\cite{Mochizuki10c}. Chirality is a universal and important concept that appears in various fields of science. Functions of biological proteins, properties of chemicals, and efficacies of drugs are inseparably related with their chirality. Methods for controlling chirality have been a long-standing challenge. For example, the asymmetric synthesis method in the field of chemistry is regarded as an epoch-making discovery. The proposal that we argue here is to control and manipulate the chirality of spins within a few picoseconds by light irradiation. The established microscopic model for the spin system in $R$MnO$_3$ enables us to perform reliable simulations for nonlinear dynamics of intensely excited Mn spins.

\begin{figure} [tb]
\begin{center}
\includegraphics[scale=1.0]{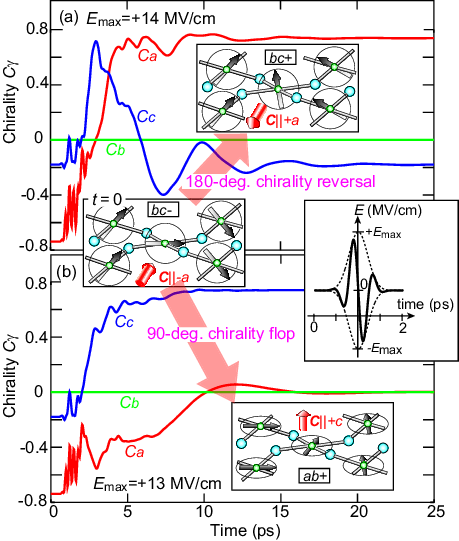}
\caption{(Color online) Theoretically simulated time profiles of $a$-, $b$-, and $c$-axis components of spin chirality after irradiation of intense laser pulse. The maximal strength of light electric field is set at $E_{\rm max}$=14 MV/cm and $E_{\rm max}$=13 MV/cm for (a) and (b), respectively. Insets show temporal waveforms of the applied laser pulses. This figure is taken and modified from Ref.~\cite{Mochizuki10d} {\copyright} 2010 American Physical Society.}
\label{Fig18}
\end{center}
\end{figure}
A sinusoidal light electric-field pulse with a Gaussian envelope is applied parallel to the $a$ axis to the ground-state $bc$-plane cycloidal spin state in TbMnO$_3$. The temporal waveform of the applied pulse is represented by,
\begin{equation}
E_\omega(t)=-E_{\rm max}\sin \omega t \exp \left[-\frac{(t-t_0)^2}{2\sigma^2}\right].
\end{equation}
The spin dynamics induced by this light pulse is simulated by solving the Landau-Lifshitz-Gilbert equation. Here the frequency $\omega$ of the light pulse was set to be 2.1 THz, which corresponds to the higher-lying electromagnon mode of TbMnO$_3$. The Gaussian half-width $\sigma$ is set to be 0.5 ps. Figure~\ref{Fig18} shows the simulated time profiles of the $a$-, $b$-, and $c$-axis components of the vector spin chirality, $\bm C$=($C_a$, $C_b$, $C_c$), which is defined by,
\begin{equation}
\bm C=\frac{1}{2N}\sum_i(\bm S_i \times \bm S_{i+\hat{x}}+\bm S_i \times \bm S_{i+\hat{y}}).
\end{equation}
When the pulse strength is $E_{\rm max}$=14 MV/cm, the $a$-axis component $C_a$, which originally had a large negative value, is reversed in sign to be positive as seen in Fig.~\ref{Fig18}(a). This indicates that the chirality reversal from the clockwise $bc$-plane cycloid to the counterclockwise $bc$-plane cycloid has occurred. In contrast, as seen in Fig.~\ref{Fig18}(b), when the pulse strength is $E_{\rm max}$=13 MV/cm, the $a$-axis component $C_a$ becomes zero, while the $c$-axis component $C_c$, which was originally almost zero, takes a large positive value. This indicates that the 90-degree flop of chirality from the $bc$-plane cycloid to the $ab$-plane cycloid has occurred. These reversal and flop processes complete quickly typically within a few picoseconds.

\begin{figure} [tb]
\begin{center}
\includegraphics[scale=1.0]{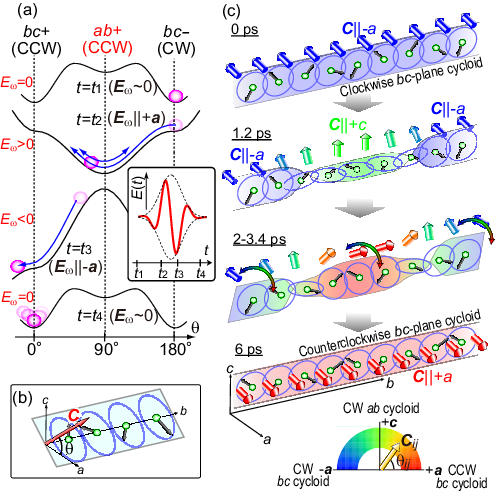}
\caption{(Color online) (a) Schematics of the temporal modulation of potential structure and the chirality reversal process induced by the light-pulse irradiation. (b) Definition of the spin-chirality angle $\theta$. (c) Formation of the dynamical spatial patterns in the transient process of the light-induced chirality reversal. This figure is taken and modified from Ref.~\cite{Mochizuki10d} {\copyright} 2010 American Physical Society.}
\label{Fig19}
\end{center}
\end{figure}
These spin-chirality reversal and flop are caused by the light-induced dynamical modulation of potential structure as well as the inertial motion of cycloidal spin plane which acquires mass due to the magnetic anisotropy. Under the light electric field $\bm E_\omega$$\parallel$$\bm a$, the ferromagnetic exchange interaction on each Mn-O-Mn bond in $R$MnO$_3$ increases or decreases as shown in Fig.~\ref{Fig15}(a). As a result, the relative angle of the spins becomes smaller (larger) on the bond where the ferromagnetic interaction becomes stronger (weaker), which results in modulation of the local spin chirality ($\bm C_{ij}$=$\bm S_i$$\times$$\bm S_j$) defined between the spins. The modulation depends on the rotation sense of the spin cycloid (clockwise or counterclockwise) or on the direction of the spin chirality $\bm C$$\propto$$\sum_{\langle i,j \rangle}$$\bm C_{ij}$. 

Because the pattern of the modulation of $\bm C_{ij}$ under $\bm E_\omega$$\parallel$$\bm a$ is different (or even opposite) depending on the rotation sense of spin cycloid, the energy of the spin cycloid becomes dependent on its rotation sense or its spin chirality in the presence of the Dzyaloshinskii-Moriya interactions coupled to $\bm S_i$$\times$$\bm S_j$. Specifically, under $\bm E_\omega$$\parallel$$+\bm a$ ($\bm E_\omega$$\parallel$$-\bm a$), the counterclockwise $ab$-plane spin cycloid has a lower (higher) energy, while the $bc$-plane spin cycloid has no energy change regardless of whether it is clockwise or counterclockwise. Consequently, as shown in Fig.~\ref{Fig19}(a), the potential structure depicted as a function of the angle $\theta$ between the spin chirality $\bm C$ and the $a$ axis [see Fig.~\ref{Fig19}(b)] dynamically changes with the sign change of the oscillating light electric field [see inset of Fig.~\ref{Fig19}(a)].

This light-induced chirality switching exhibits interesting phase-transition dynamics accompanied by formation of dynamical spatial patterns. Specifically, dynamical stripes of the chirality domains are formed as schematically shown in Fig.~\ref{Fig19}(c). When the clockwise $bc$-plane spin cycloid is irradiated with a light pulse, stripe domains of chirality are generated, and each domain oscillates in a phase opposite to the neighboring domains and eventually settles into the counterclockwise $bc$-plane spin cycloid.

In order to realize this chirality switching, the frequency of light must be fixed at the resonance frequency of the higher-lying electromagnon mode to excite the mode efficiently. A strong light electric field ($E_{\rm max}$$\sim$10 MV/cm in the present case) is also required, but stronger is not always better. As can be seen in Fig~\ref{Fig19}(a), the chirality oscillation at a potential bottom under $\bm E_\omega$$\parallel$$+\bm a$ at time $t$=$t_2$ and the sign reversal of the light electric field at time $t$=$t_3$ must be well synchronized such that the chirality falls into the opposite potential bottom. If the timing is well synchronized, the 90-degree flop to the $ab$-plane spin cycloid might also be realized~\cite{Mochizuki11b}. We expect that the threshold intensity of the light electric field can be reduced with a longer pulse duration. Switching can also be achieved with a weaker light electric field by taking advantage of assist from thermal fluctuations or by selecting a material close to the phase boundary between cycloidal magnetic states with different cycloidal planes. Simulations have also shown that it is possible to switch among four different chirality states (clockwise and counterclockwise $ab$-/$bc$-plane cycloidal spin states) at will by changing the sign, shape, length, and intensity of light pulses.

\section{Conclusion}
In this article, we have discussed the material properties and the physical phenomena of the rare-earth perovskite manganites $R$MnO$_3$ by particularly focusing on the multiferroic nature and the magnetoelectric phenomena. The multiferroic phases with ferroelectricity induced by nontrivial magnetic orders in $R$MnO$_3$ are observed in the compounds with moderate sized $R$ ions (e.g., Tb, Dy, Eu$_{1-x}$Y$_x$, etc) and the compounds with small sized $R$ ions (e.g., Ho, Y, $\cdots$, Lu). The ferroelectric polarization is induced by the cycloidal spin order via the spin current mechanism or the inverse Dzyaloshinskii-Moriya mechanism associated with outer-products of adjacent Mn spins $\bm S_i \times \bm S_j$ in the former system, while in the latter system, the ferroelectric polarization is induced by the $E$-type antiferromagnetic order with four-fold up-up-down-down spin alignment via the symmetric magnetostriction mechanism associated with inner-products of adjacent Mn spins $\bm S_i \cdot \bm S_j$. Although we have mainly discussed the former system with cycloidal spin orders because of the limited pages, both spin orders are stabilized by the frustrated exchange interactions in the $ab$ plane caused by the significant GdFeO$_3$-type lattice distortion and the special electronic structure of Mn$^{3+}$ ions with $t_{2g}^3$$e_g^1$ electron configuration. The magnetic frustration effectively suppresses an energy scale of the exchange interactions and inevitably enhances relative importance of other weak interactions and small anisotropies. In this article, by constructing a microscopic model of $R$MnO$_3$ by considering these interactions and anisotropies, we have demonstrated that keen competitions among the Dzyaloshinskii-Moriya interactions, the single-ion magnetic anisotropies, and the spin-phonon coupling exist behind the rich magnetoelectric phenomena in $R$MnO$_3$. These phenomena as well as their physical mechanisms encompass universal physics expected in many other multiferroic materials. The research on multiferroics is still on-going. We hope that this article will be of some help for future progress of the research.

\section{Acknowledgment}
Most of the theoretical studies discussed in this article were conducted in connection with the ERATO multiferroic project,  Japan Science and Technology Agency, led by Prof. Yoshinori Tokura. I would like to express my cordial gratitude to him for giving me a precious opportunity to work on the multiferroics as a member of the project with many diligent professors, researchers, and colleagues.

\end{document}